\documentclass[aps,prl,superscriptaddress]{revtex4}

\usepackage[breaklinks,colorlinks,urlcolor=blue,citecolor=blue,pdfa=true,backref=page]{hyperref}

\usepackage{amsmath}  
\usepackage{amsfonts} 
\usepackage{graphicx} 
\usepackage{url}

\usepackage[breaklinks,colorlinks,citecolor=blue,pdfa=true,backref=page]{hyperref}

\begin{document}

\title{A Bose Horn Antenna Radio Telescope (BHARAT) design  for 21 cm hydrogen line experiments for  radio astronomy teaching}

\author{Ashish A. Mhaske}
\email{ashishm@iucaa.in} 
\affiliation{Inter-University Centre for Astronomy and Astrophysics, Post Bag 4, Ganeshkhind, Savitribai Phule Pune University Campus, Pune 411007, India}

\author{Joydeep Bagchi}
\email{joydeep.bagchi@christuniversity.in}
\affiliation{Department of Physics \& Electronics, CHRIST (Deemed to be University), Hosur Road, Bengaluru 560029, India}
\affiliation{Inter-University Centre for Astronomy and Astrophysics, Post Bag 4, Ganeshkhind, Savitribai Phule Pune University Campus, Pune 411007, India}

\author{Bhal Chandra Joshi}
\email{bcj@ncra.tifr.res.in}
\affiliation{National Centre for Radio Astrophysics - Tata Institute of Fundamental Research, Pune University Campus, Post Bag 3, Ganeshkhind Pune 411007 India}

\author{Joe Jacob}
\email{drjoephysics@gmail.com}
\affiliation{Newman College, Thodupuzha, Kerala 685585, India }

\author{Paul K.T.}
\email{paul.kt@christuniversity.in}
\affiliation{Department of Physics \& Electronics, CHRIST (Deemed to be University), Hosur Road, Bengaluru, 560029, India}

\date{\today}

\begin{abstract}
We have designed a low-cost radio telescope system named the Bose Horn Antenna Radio Telescope (BHARAT) to detect the 21 cm hydrogen line emission from our Galaxy. The system is being used at the Radio Physics Laboratory (RPL),\cite{RPL} Inter-University Centre for Astronomy and Astrophysics (IUCAA), India, for laboratory sessions and training students and teachers. It is also a part of the laboratory curriculum at several  universities and colleges. Here, we present the design of a highly efficient, easy to build, and cost-effective dual-mode conical horn used as a radio telescope and describe the calibration procedure. We also present some model observation data acquired using the telescope for facilitating easy incorporation of this experiment in the laboratory curriculum of undergraduate or post-graduate programs. We have named the antenna after {\it Acharya}\footnote{Teacher or an influential mentor.}  Jagadish Chandra Bose, honoring a  pioneer in radio-wave science and 
an outstanding teacher, who inspired several world renowned
scientists.

\end{abstract}

\maketitle

\section{Introduction} 
Radio frequency antennas were initially developed for communication purposes. 
 Forerunners of modern radio antennas can be found in the initial experiments conducted by pioneers like  Heinrich Hertz, Oliver Lodge, G.F. Fitzgerald, Oliver Heaviside, and Jagadish  Chandra Bose. The experiments conducted by these people were motivated by the stupendous discovery of James Clerk Maxwell, which asserted that visible light is simply a small section of the electromagnetic spectrum, where all electromagnetic waves travel at the speed of light. Maxwell's discovery suggested that, just as visible light, radio waves should exhibit all of the properties of electromagnetic waves. 

 The subsequent detection of radio frequency radiations from outer space was quite serendipitous.\cite{Jansky, smith, ra_book, Condon} However, radio astronomy has remained elusive to students and educators for a long time due to the apparent complexity and lack of readily available hands-on experimental components. Recent advancements in electronics and the capabilities of modern computers have facilitated the miniaturization of electronic components and brought with it a reduction in the complexity of instrumentation. Hence, with the current increase in demand for such devices, high-frequency radio receiver components are now available commercially at an affordable price. These components have served to simplify experimental astronomy setups and make them easily accessible to universities, colleges, and the general public. 
 
 The 21 cm hydrogen line measurement setup that we present here is designed to instruct students in the instrumentation, observational techniques, and data analysis methods used in radio astronomy. A major component of the setup is the antenna, which students can construct by themselves or with the help of a lab technician. The dual-mode horn antenna used in our system has low sidelobes to reduce the noise picked up from the ground and surrounding structures.

\section{21-cm Line of Neutral Hydrogen}

\begin{figure*}[ht!]
\centering
\includegraphics[scale=0.2]{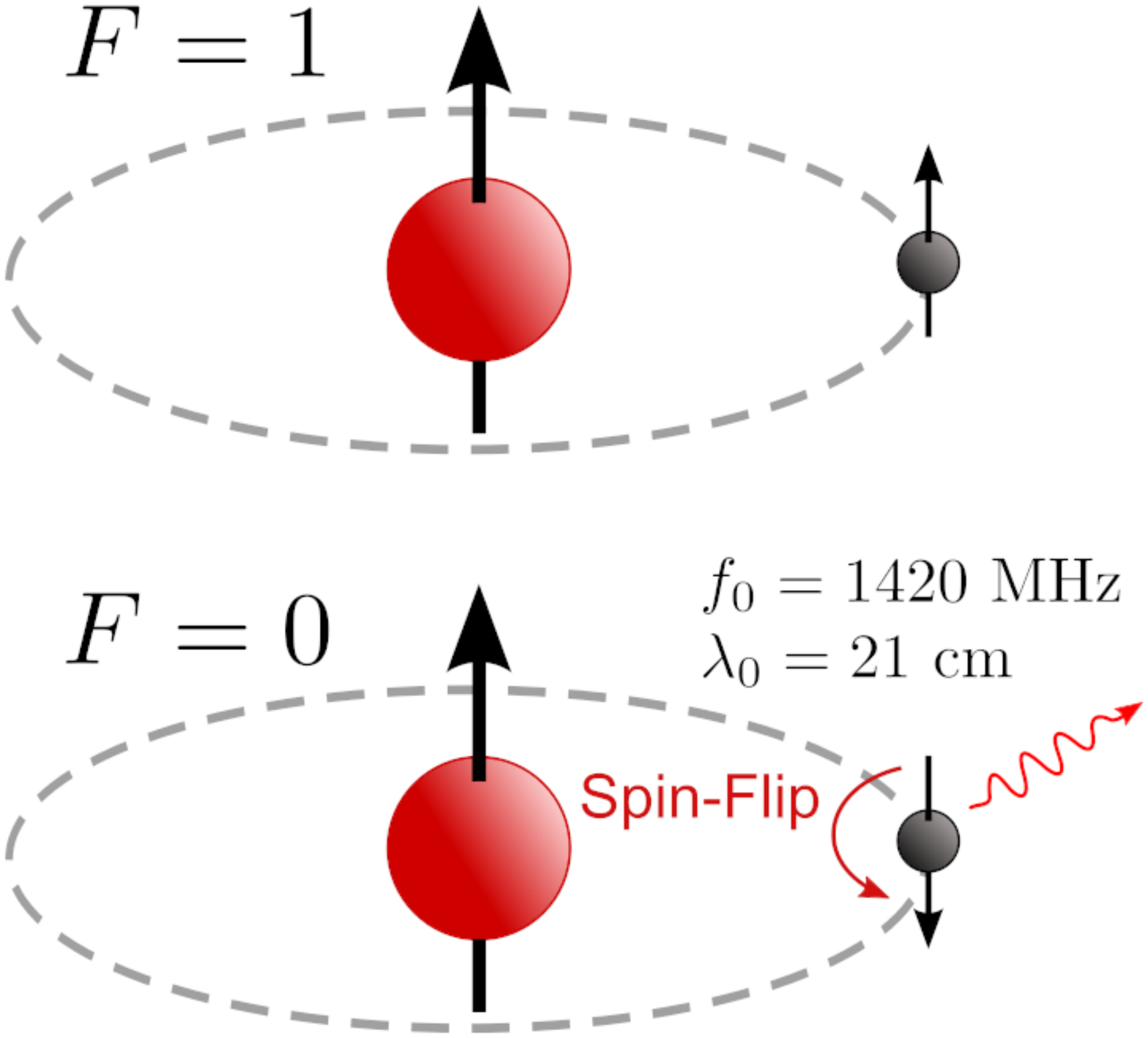}
\includegraphics[scale=0.36]{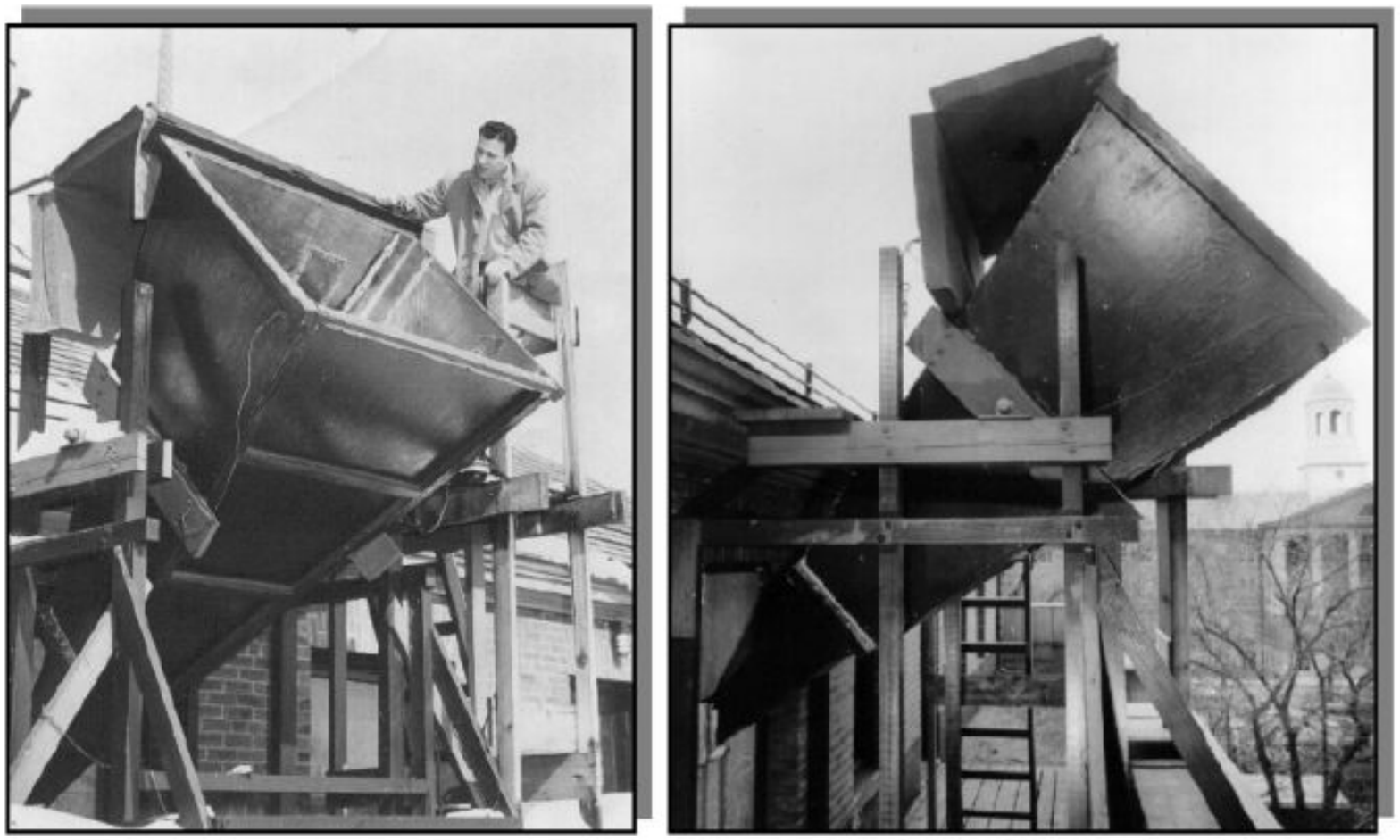}
\caption{(Color online) Left figure: In a  neutral hydrogen atom, the ground state energy level  is split by an  extremely small (hyperfine) energy difference of about $5.87$ micro electron volt. In the higher energy (excited) state, the proton (red ball) and electron (black ball) spins are parallel (F = 1 state), and  the lower energy state has  anti-parallel spins (F = 0 state). Normally most of the atoms would reside in ground state and a small fraction in excited state. When a spin-flip transition happens from higher to the lower energy state, radio radiation at a frequency of 1420 MHz (5.87 $\mu$eV energy) is emitted; the corresponding wavelength is 21.1 cm. This is a highly simplified picture of the origin of the 21-cm  hydrogen line (Figure Credit: Wikipedia). Right Figure: This horn antenna, mounted on a fourth floor window of the Lyman Physics Laboratory at Harvard University (now at NRAO, Green Bank, Charlottesville, USA), was used to  first detect the radio radiation from the neutral hydrogen gas in the Milky Way on March 25, 1951 (Figure Credit: National Radio Astronomy Observatory.)}
\label{spin-flip}
\end{figure*}

In 1945, H.C. Van de Hulst predicted the 21 cm line radiation from neutral hydrogen.\cite{Hulst_21} After several unsuccessful attempts, on  March 25, 1951, Harold ``Doc" Ewen and his thesis advisor Edward M. Purcell made the first detection of this hydrogen line at Harvard University using a horn-shaped radio antenna.\cite{Ewen} To make this observation, these researchers stuck their antenna out of a window on the fourth floor of the Lyman Physics Laboratory (Fig.~\ref{spin-flip}). About 17 days later, the observation was repeated by C.A Muller and J.H. Oort in the Netherlands.\cite{Muller_Oort} These  pioneering  works are the foundation of further discoveries that allow us to see the universe in a way never possible before. After the 21 cm line became measurable, Van de Hulst, with Jan Oort and C.A. Muller, used radio astronomy to map out the neutral hydrogen in our Galaxy, which first revealed its majestic spiral structure.  An excellent account of the first detection of the 21 cm hydrogen line is given in Ref.~\onlinecite{stephan}. 

In a simplified picture, the hydrogen atom has an electron and a proton, both of which behave like spinning magnetized tops; these tops can have parallel or anti-parallel spins, with the former configuration having a slightly higher energy than the latter. When a hydrogen atom transitions from a parallel spin to an anti-parallel spin configuration, it emits electromagnetic radiation corresponding to  a frequency of 1420.4 MHz or, equivalently, a wavelength of 21.1 cm (Fig.~\ref{spin-flip}). In the quantum picture, the spin of the electron or proton does not represent a classical spinning charge sphere literally. Rather, it describes  the behavior of  particle's quantum mechanical angular momentum  (called \emph{spin}) and differs from the classical analogy. A  more precise value for the spin-flip transition measured in  laboratory is  21.106114054160(30) cm or 1420405751.768(2) Hz in vacuum.

The probability of spontaneous emission of 21 cm radiation from a hydrogen atom is, on average, about one in 11 million years. However, the abundance of hydrogen atoms in the interstellar medium of our Galaxy makes it possible to detect this line in emission. Due to the long wavelength, the interstellar medium gives minimal attenuation to this line. 

Hence, the galactic rotation curve and the structure of the Milky Way galaxy, as well as nearby galaxies, can be studied using the 21 cm line. Van de Hulst \emph{et al.} made the early estimate of the galactic rotation curve from neutral hydrogen line observations\cite{hulst} and produced a map of the outer galactic structure. John Findlay built the Little Big Horn in Green Bank, which was used for absolute calibration of the supernova remnant Cassiopeia A at 21 cm.\cite{little_big_horn} A detailed account of the calibration can be found in Ref.~\onlinecite{Findlay}. 
 
The rotation curve of the Milky Way and  other spiral galaxies derived from  21 cm hydrogen line observations provide strong evidence for the still mysterious dark-matter.\cite{Corbelli} In recent years, the quest for detecting the signature of the Epoch of Reionization has brought this line (red-shifted to low frequency by the expansion of the Universe) to the forefront of cosmology.\cite{saras, edges} These developments clearly show the rich astrophysics that can be done by studying this important spectral line. 

\section{Jagadish Chandra Bose (1858 - 1937): a forgotten pioneer of radio wave science}

\begin{figure*}[ht!]
\centering
\includegraphics[scale=0.25]{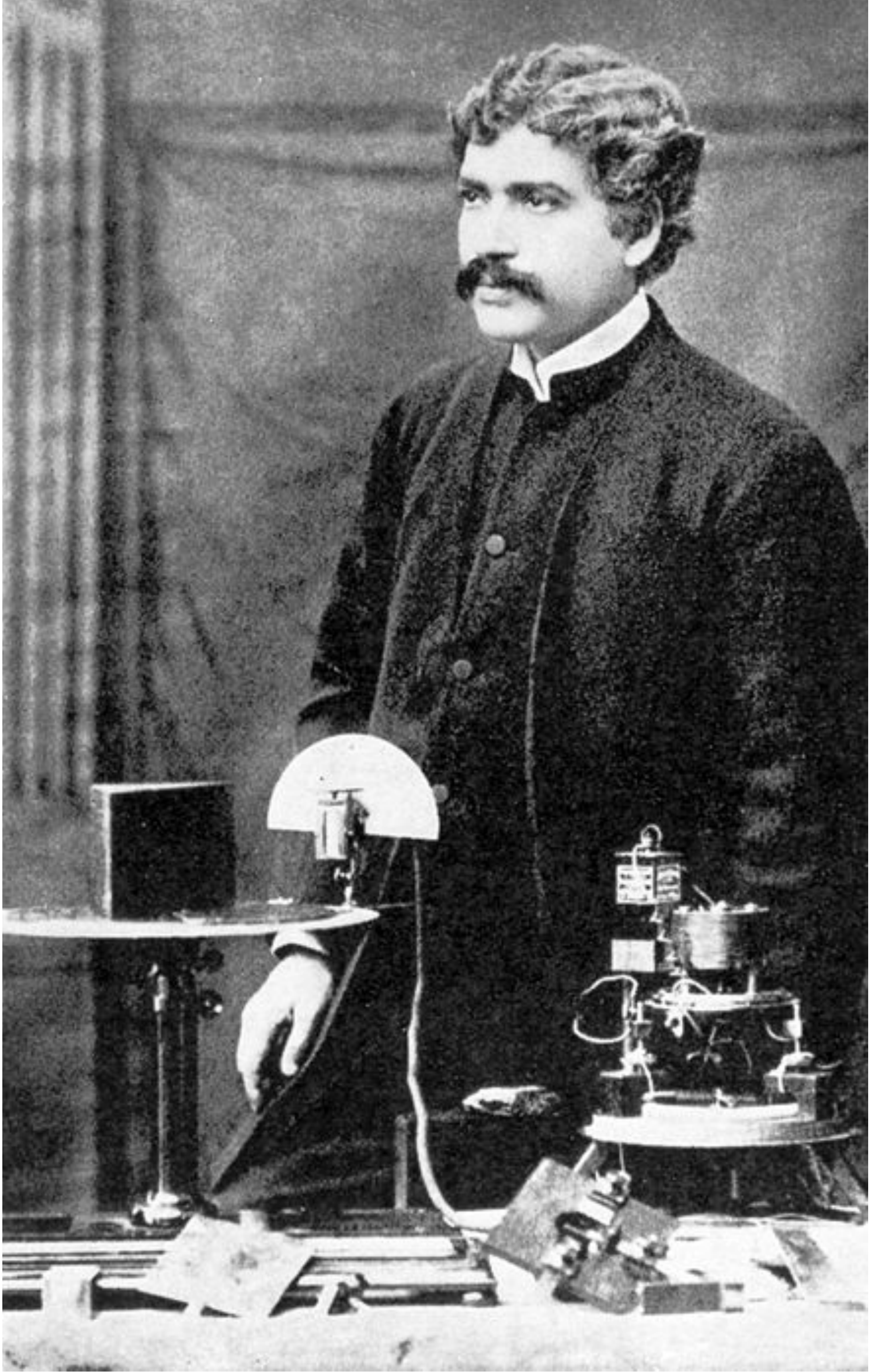}
\includegraphics[scale=0.51]{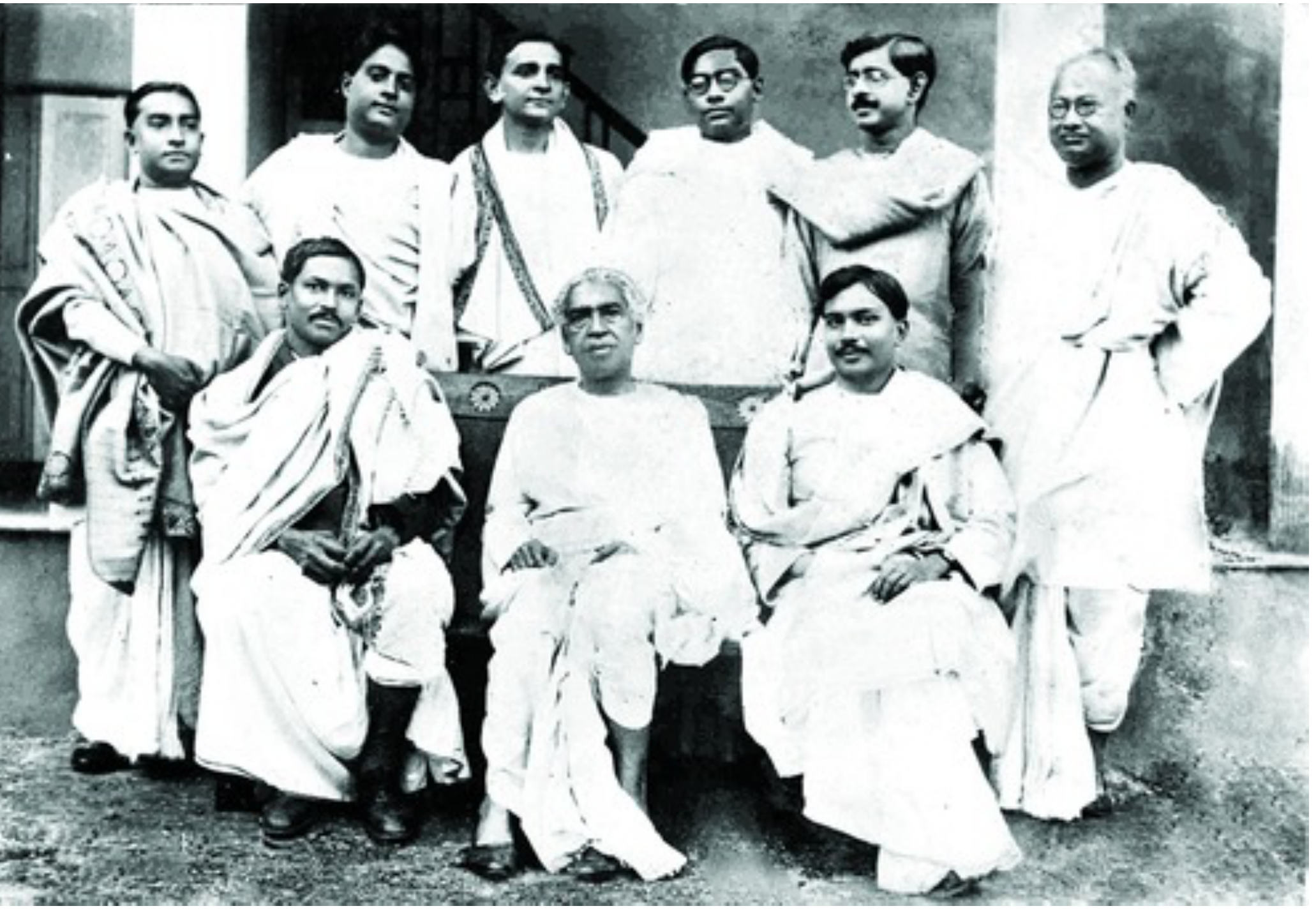}
\caption{Left: Sir J. C. Bose at Royal Institution London in 1897, lecturing on his radio wave communication experiments.\cite{bose_pic} Bose used an early version of the modern pyramidal horn antenna and waveguide transmitter in this demonstration.
Right: Bose in his later years with some of his students and colleagues at the
Bose  Institute  Calcutta. In this rare photograph are seen Jagadish Chandra Bose (seated, centre),  physicist Satyendranath Bose (of Boson and Bose-Einstein statistics fame - standing, second from left), astrophysicist Meghnad Saha (formulated Saha equation of thermal ionization - seated, left), physicist  Debendra Mohan Bose (Cosmic ray pioneer,  first  photographed tracks of Mesons ($\mu, \pi$), whose discovery was later announced by C.F. Powell - standing, third from left) 
and other eminent persons. Photograph from the archive of Saha Institute of Nuclear Physics, Kolkata. 
}
\label{Bose}
\end{figure*}

Sir Jagadish Chandra Bose was an Indian scientist of international repute whose work laid the foundation for millimeter-wave radio science and plant physiology.\cite{Bose_Pubs} Not only he was a  pioneering scientist, he was also an extraordinary teacher, science writer, and accomplished poet. He  produced a galaxy of outstanding students who  attained worldwide fame (Fig. \ref{Bose}). Bose's experiments in microwave optics were specifically aimed at understanding fundamental physics rather than developing radio into a commercial communication medium. He performed his pioneering experiments around the same period (from late 1894 onward) that Guglielmo Marconi was making breakthroughs on a radio system specifically designed for  wireless telegraphy. Bose was the first to use a doped semiconductor junction to detect radio waves and he also invented various now commonplace microwave components, including the first horn antenna and waveguide.\cite{bose1, bose2} Bose's experimentsnhelped form the foundation of short wavelength ($\sim$100 GHz) radio wave technology, which was fully developed later on.\cite{bose3}

In 1895, he demonstrated the generation and reception of radio waves at the Calcutta Town Hall, about two year before the historic public demonstration by Marconi at Salisbury Plain in 1897. By 1902, Marconi attained worldwide fame by sending a wireless telegraph message from Poldhu, Cornwall in southwest England to Glace Bay, in Nova Scotia, Canada, a distance of more than 2,100 miles across the Atlantic. 
The mercury `coherer' type diode  detector used by Marconi in his experiments was actually modified  from an original 1899 invention by Bose.\cite{coherer} In 1909, Marconi shared the Nobel Prize for physics with Karl Ferdinand Braun of Germany.  In India, during the later phase of his career, Bose's interest shifted to plant life. Bose made a number of  pioneering discoveries in plant physiology and  proved, using  highly sensitive, ingenious instruments he built himself, the parallelism between animal and plant tissues. Unfortunately, because of the colonial mentality of his time, along with the fact that he refused to patent his inventions, Bose is rarely given his deserved recognition outside India. In a recent befitting tribute to Bose's pioneering contributions, his laboratory in Calcutta (now Kolkata) received  the `IEEE Milestone' award.\cite{ieee} Thanks to his breakthrough inventions, we can  communicate easily, and uniquely probe the universe using radio waves. As a rich tribute to him, we have dedicated our horn-antenna-based radio telescope design to  {\it Acharya} Jagadish Chandra Bose, honoring a pioneer in radio-wave science and an extraordinarily inspiring teacher.  

\section{Instrument}
\subsection{Dual-mode horn}

\begin{figure}[h!]
\centering
\includegraphics[scale=0.4]{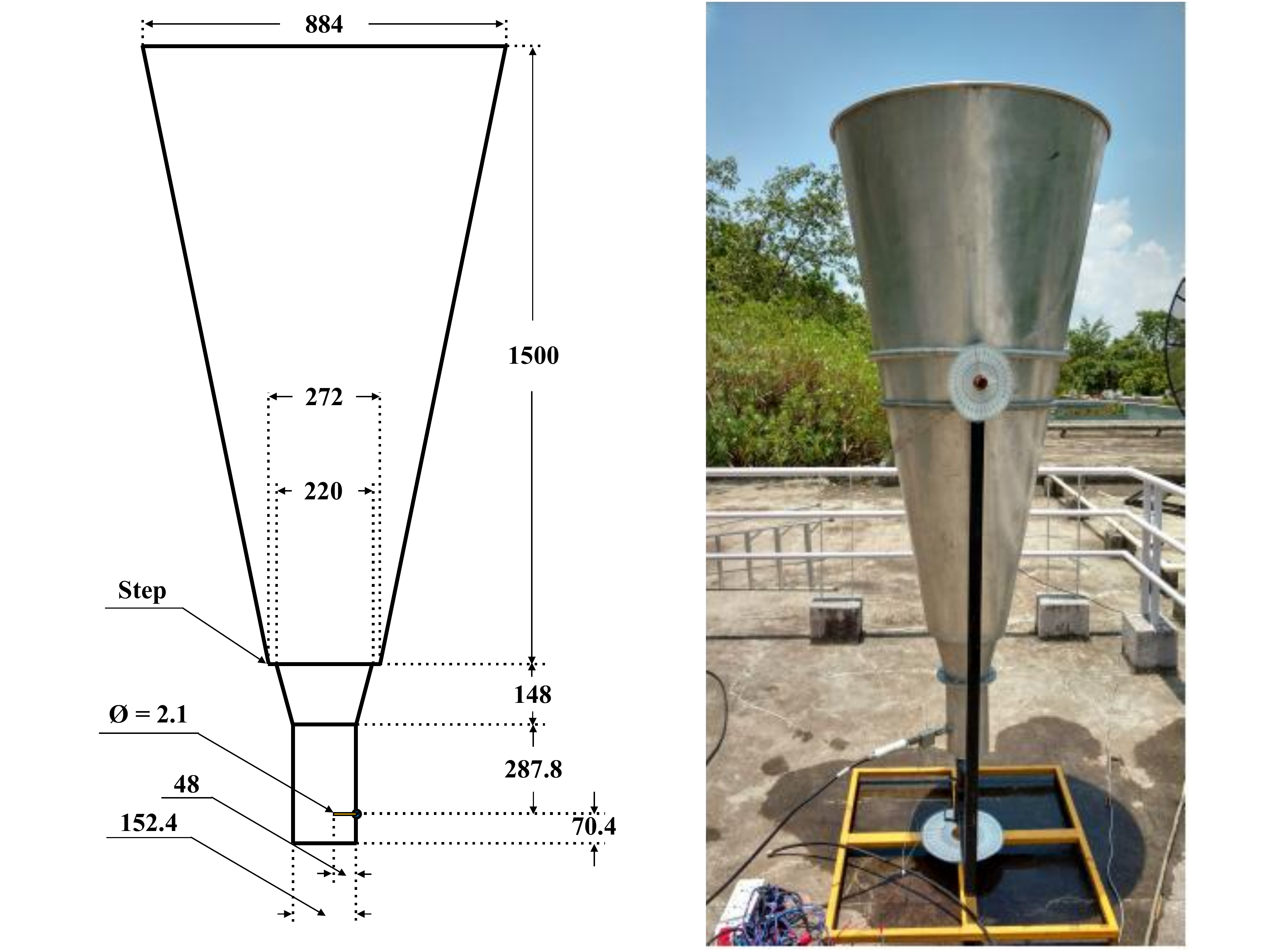}
\caption{Dimensions (in mm) of the dual-mode horn are shown on the left. Note the step discontinuity in the horn flare, which generates a second mode. The horn is fed using an N-type coaxial connector. On the right is the constructed horn fixed on an altitude-azimuth mount. Scales are added for pointing the antenna in a particular direction. A mobile phone running an appropriate application for showing the azimuth and elevation values can also be alternatively fixed on the system for pointing purposes.}
\label{dual}
\end{figure}

Horn antennas are closely linked to the developments in radio astronomy. The first detection of 21 cm hydrogen line,\cite{Ewen} the discovery  of Cosmic Microwave Background (CMB),\cite{CMB} and the design of modern radio telescopes show that horns have been widely used, and are presently being used, in astronomy due to their superior antenna pattern and  wider frequency characteristics. As the design is simple, a horn is easier to construct than most of the other antennas with similar radiation characteristics. P. D. Potter developed a dual-mode horn with significantly low sidelobes compared to a simple horn.\cite{potter} A modified design based on Potter horn, and proposed by Bailey,\cite{bailey} is adopted for the present antenna.

In order to avoid the possible radiation contamination from the surroundings when this telescope is installed in educational institutions, it is necessary to give special care in designing an antenna with low sidelobes and high gain. So we choose a dual-mode horn design satisfying these criteria. A waveguide can be excited in transverse electric (TE) or transverse magnetic (TM) modes. The horn being a flared extension of a waveguide, one can imagine it as a cavity that can be excited in different modes of electromagnetic waves. The horn discussed here is initially excited in the fundamental mode, i.e., TE11 mode. Then, a second mode, i.e., TM11 mode, is generated by introducing a step discontinuity in the cavity, as shown in Fig. \ref{dual}. Hence, this type of horn is also known as a dual-mode horn. The purpose of generating the second mode is to reduce the sidelobes in the far field. The far-field pattern of an antenna is given by the Fourier transform of the electric field distribution on the aperture.\cite{nrao} If the electric field on the aperture is tapered, then the sidelobes in the far-field will be reduced. This tapering is achieved by the superposition of the two modes mentioned above at the aperture, which cancel each other at the edge of the aperture.\cite{potter}

\begin{figure}[h!]
\centering
\includegraphics[scale=0.7]{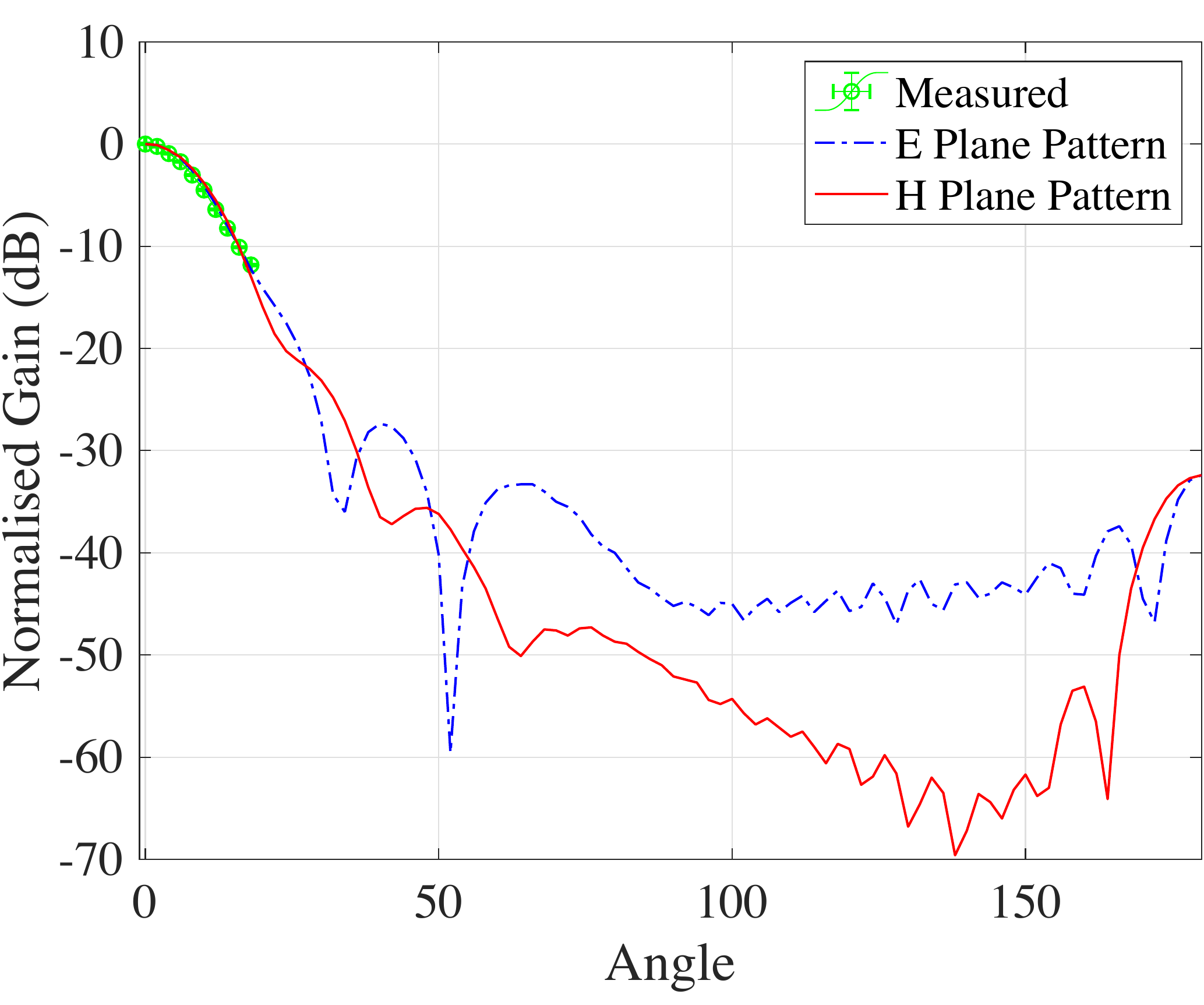}
\caption{Simulated and measured radiation patterns of the proposed horn. The worst sidelobe level is $-27$ dB in E-plane (blue curve). The measurement could only be done in E-plane due to limitations of the setup. The measured pattern is limited to $-12$ dB due to the sensitivity limitation of the setup.}
\label{dual_rad}
\end{figure}

Electromagnetic simulations for the antenna are done using Altair FEKO,\cite{FEKO} which is a commercial computational electromagnetic software program. The waveguide termination and waveguide-to-coax transition are also optimized in FEKO. The simulated and measured radiation patterns of the designed antenna are presented in Fig.~\ref{dual_rad}. To measure the radiation pattern, a transmitting antenna is placed in the far-field region. The horn is the receiving antenna which is rotated to obtain the radiation pattern. The following equation gives an estimate of the distance $R$ where the far-field region starts: 

\begin{equation}\label{eq:farfield}
R \geq \frac{2D^2}{\lambda}
\end{equation}

where $D$ is the aperture diameter, and $\lambda$ is the operating wavelength. The radiation pattern is measured to determine the full-width half maximum (FWHM) of the main lobe of the radiation pattern of the antenna. Main lobe is the lobe containing peak of the radiation pattern. FWHM is the angular width between points on the main lobe where the gain of the main lobe falls down by half (3 dB) from the peak value. 
As a rough guide, the far-field distance $R$ is $> 10$ m for a horn antenna of 1 m  aperture
radiating at 1420 MHz. In the far-field  zone (also called the \emph{Fraunhofer radiation-zone})  of an antenna, (a) the electric and magnetic fields show   electric dipole like  predictable field patterns, (b) the  radiated power decreases as the inverse square of distance, and  (c) the electromagnetic field is radiative in nature. This means that, in the far-field region, all of the electromagnetic field energy of the antenna  effectively escapes to infinite distance.

The experimental results are in good agreement with the simulations. The FWHM of the antenna is 16.5$^{\circ}$, and the gain is 20.6 dBi. The effective aperture area $A_e$ for a gain of 20.6 dBi can be calculated using the following equation:\cite{Condon}

\begin{equation}\label{eq:Ae}
A_e = \frac{\lambda^2 G}{4\pi} = 0.407 \,\rm{m}^2
\end{equation}

The aperture efficiency is then defined as,

\begin{equation}\label{eq:n}
\eta = \frac{\text{effective aperture area}}{\text{physical aperture area}} = 66.3\%
\end{equation}

The aperture efficiency is expected to be slightly lower than that of standard conical horns because the effective area is further reduced due to the cancellation of electric fields resulting from the superposition of the two modes (TE11 and TM11) at the edge of the aperture. This results in a more tapered electric field distribution on the aperture, which in turn gives symmetric beam and lower sidelobes in the far-field. The sensitivity of the antenna is another important parameter and is given by the following relation:\cite{Condon}

\begin{equation}\label{eq:sensitivity}
\zeta = \frac{A_e}{2k} = 1.47 \times 10^{-4} \,\rm{K/Jy}
\end{equation}
Let us compare this value to that calculated for the 100-meter diameter parabolic antenna at the Effelsberg radio observatory. The Effelsberg antenna has a beam-width $\theta $ of 587 arc seconds at 21 cm wavelength,\cite{Effelsberg_resolution} which gives us the effective area using\cite{Condon}

\begin{equation}\label{eq:effective diameter}
A_e = \frac{\lambda^{2}}{\theta^{2}} = 5.5 \times 10^{3} m^{2}
\end{equation}

It then follows that the aperture efficiency of this antenna is about $ 70\%$ and the sensitivity is 1.9 K/Jy, which can be calculated from Eq.~(\ref{eq:n}) and Eq.~(\ref{eq:sensitivity}), respectively.

Measuring the voltage standing wave ratio (VSWR) provides information about impedance matching between the antenna and the front-end electronics. The first component connected to the antenna is a low noise amplifier (LNA) with an input impedance of 50 $\Omega$. Since this is an industry standard, we try to design the antenna port to match this impedance. An impedance mismatch will produce standing waves, which are formed due to the incident and reflected waves at the port, and will reduce the instrument sensitivity. VSWR is defined as follows:

\begin{equation}\label{eq:VSWR}
\rm{VSWR} = \frac{1+\Gamma}{1-\Gamma}
\end{equation}

\begin{equation}\label{eq:tau}
\Gamma = \frac{Z_l-Z_0}{Z_l+Z_0}
\end{equation}

where $\Gamma$ is the reflection coefficient, $Z_l$ is the impedance of the antenna port, and $Z_0$ is the characteristic impedance that is kept at 50 $\Omega$. The ideal value for VSWR is 1, and practically it can only be greater than 1.
The simulated VSWR of the antenna, which is found to be 1.18, is close to the measured value of 1.11, ensuring that there is an excellent impedance match to the 50 $\Omega$ load, i.e., LNA.

The antenna is mounted on an altitude-azimuth mount and is steerable by hand. A mechanism is provided to lock the antenna when pointing towards various directions during the observation. Since the antenna is not designed for tracking the sources, the integration time is kept as 4 s to avoid a significant change in the source position in the sky and to get a good signal-to-noise ratio (SNR).

\subsection{Electronics}

\begin{figure}[h!]
\centering
\includegraphics[scale=0.4]{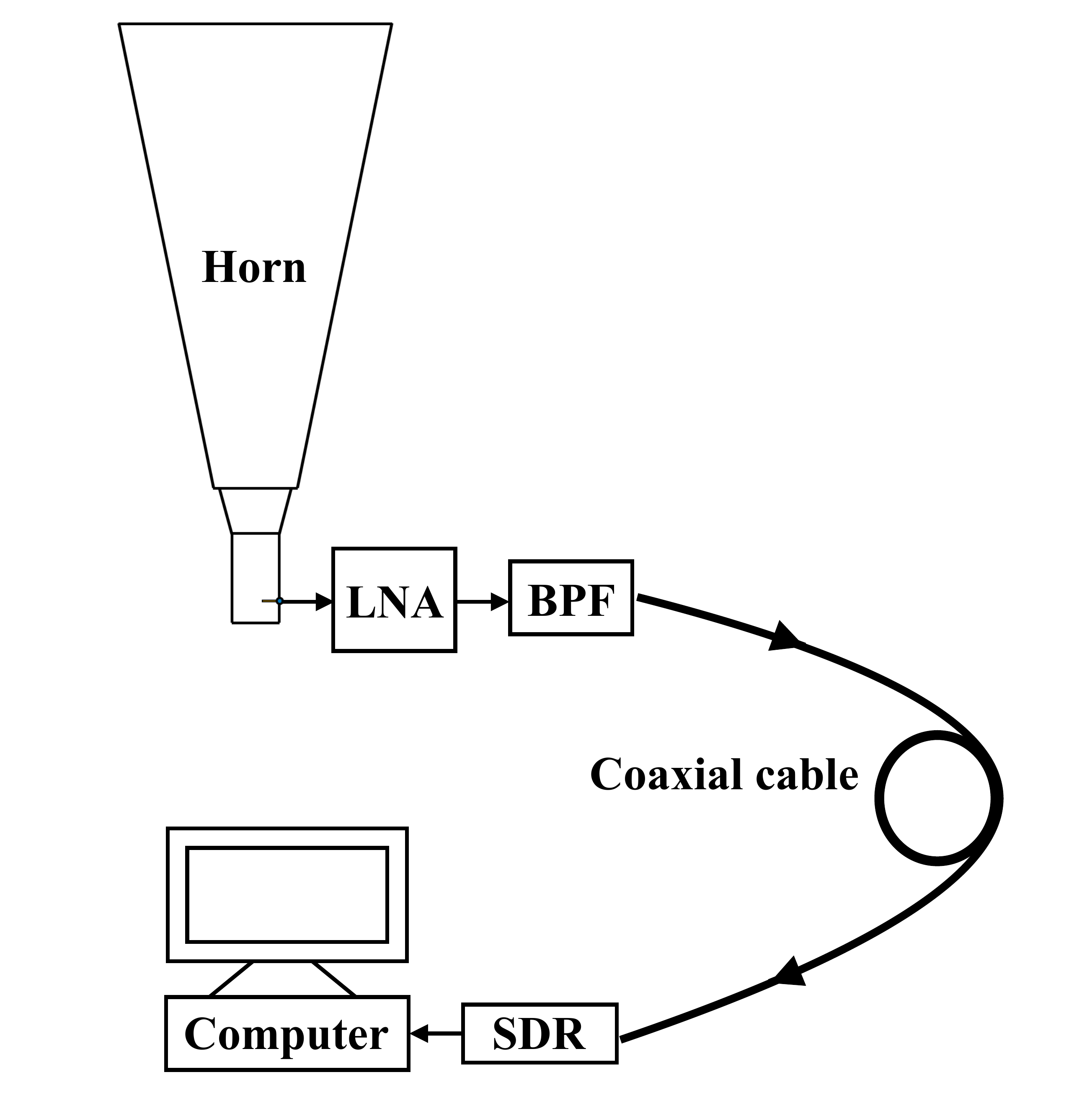}
\caption{Setup of the experiment. The low noise amplifier (LNA) and band pass filter (BPF) form the front-end electronics. A long, low-loss coaxial cable is used to connect the front-end to the back-end. Use of a long cable is recommended in order to avoid possible RFI generated by the computer. In the current setup, an LMR 400 cable is used. Other low-cost RF cables, e.g., RG58,  with 50 $\Omega$ characteristic impedance, can also be used.}
\label{setup}
\end{figure}
Handling radio frequency electronics at 1.42 GHz can be challenging for students. Hence, off-the-shelf components were preferred in our design to provide a plug-and-play telescope system, i.e., students do not have to create and assemble their own electronic circuits from scratch. Thus, the resulting electronics setup is fairly simple. The front-end consists of an LNA and a filter, which were procured from Radio Astronomy Supplies.(RAS)\cite{ras} See Table~\ref{table} for more details. The LNA has a 37 dB gain and 0.29 dB noise figure, which translates to a 20 K noise temperature. The filter bandwidth is 50 MHz with a 2 dB loss in the passband. The signal from the filter is fed directly to the receiver via a coaxial cable. The receiver used for the experiment is an inexpensive Software Defined Radio (SDR) dongle,\cite{rtl_sdr} which has become a popular choice due to its low cost, ease of availability, and wide applications. The dongle is easy to set up and has the advantage that there is a large community dedicated to experimenting with SDRs. A brief summ
ary of the dongle can be found in Ref.~\onlinecite{rtl_sdr_datasheet}. The tuner IC in the dongle, i.e., R820T2, takes care of heterodyning and downconverting the rf signal.\cite{tuner_IC} It allows selection of rf frequency and tuner gain, with tuning from 24 MHz to 1700 MHz and variable sampling rate up to 2.56 MS/s. The device uses a 28.8 MHz temperature compensated crystal oscillator (TCXO). Frequency stability of the dongle was checked by feeding a signal from a frequency generator with GPS disciplined oscillators and also by comparing observations of the S8 region as seen in Fig. \ref{S8}. No measurable frequency drift was observed throughout the observation session. Details of other stability tests can be found in Ref.~\onlinecite{stability}.
The software used for data acquisition is RTL SDR Scanner, available at EartoOak.\cite{scanner}.
 Table~\ref{table} lists the costs for two versions of the radio telescope setup, one with a little higher budget than the other. The measurements presented here are done using the dual-mode horn setup with RAS LNAs and filters. See Appendix B for the details of the pyramidal horn setup.
 
\begin{table}[h!]
\centering
\caption{Cost (in USD) and comparison of the two setups.}
\begin{ruledtabular}
\begin{tabular}{l c r l c r }
Component & quantity & cost & Component & Quantity & Cost \\
\hline
Dual Mode Horn & 1 & \$120 & Pyramidal Horn & 1 & \$100 \\
RAS LNA\cite{ras} & 1 & \$230 & Minicircuits LNA\cite{mini_bpf} & 2 & \$108  \\
RAS Filter\cite{ras} & 1 & \$150 & Minicircuits Filter\cite{mini_amp} & 1 & \$56 \\
Cables and connectors & - & \$50 & Cables and connectors & - & \$50  \\
SDR dongle\cite{rtl_sdr} & 1 & \$22 & SDR dongle\cite{rtl_sdr} & 1 & \$22  \\
\hline
Total & & \$572 & & & \$336
\end{tabular}
\end{ruledtabular}
\label{table}
\end{table}
\section{Calibration and Measurement}

The spectral plot acquired using our system yields power in arbitrary units, which then needs to be converted to a temperature scale using calibration sources. Two sources with known temperatures are required to determine the noise equivalent system temperature. One source can be the sky ($T_{sky}$), which is be assumed to be at 5 K (this includes CMB and atmospheric contributions),\cite{vine} and the other source can be a nearby wall with a temperature denoted as $T_{ref}$, which assumed to be at 300 K. These are the off-source and on-source measurements, respectively. Since neutral atomic hydrogen (also called HI, I is Roman numeral hence pronounced H one and represents the ionisation state of the atom) emission is present in all directions in the sky, we choose a part of the sky with a high elevation and relatively low galactic HI emission for the calibration. We consider only the baseline, i.e., the measurement at the frequencies where the emission is absent, as the off-source measurement. Basically, we calibrate the nearby frequencies where the hydrogen line is not present and extrapolate the calibration to the frequencies where the line is present. With these two measurements, we can find the system temperature and the brightness temperature of the line. 
If the antenna is kept in an environment with temperature $T$ and terminated with a matched load, it will be in equilibrium with the surrounding temperature. The power absorbed by the load via the antenna is then proportional to the temperature of the surroundings. This is given by $P = kT$, where $P$ is the spectral power in units W/Hz and $k$ is the Boltzmann's constant. Hence, we can convert the power measured by the receiver into a temperature. Since the receiver is not calibrated, the power measured is not absolute.

But we can define the following equation.
\begin{equation}\label{eq:PonPoff}
\frac{P_{on}}{P_{off}} \cong \frac{T_{on}}{T_{off}} 
\end{equation}
$P_{on}$ is the power received when the antenna points to the wall, and $P_{off}$ is the power when the antenna points to the sky; these are the measured quantities. $T_{on}$ and $T_{off}$ are corresponding temperatures related by 
\begin{equation}\label{eq:TonToff}
\frac{T_{on}}{T_{off}} = \frac{T_{ref} + T_{r}}{T_{sky} + T_{r}}  
\end{equation}
where $T_{ref}$ is the temperature of the wall, which is assumed to be 300 K, $T_{r}$ is the radiometer temperature, and $T_{sky}$ is the sky temperature, which is 5 K. Since we know $P_{on}$ and $P_{off}$,  Eqs. (\ref{eq:PonPoff}) and (\ref{eq:TonToff}) give
\begin{equation}\label{eq:Tr}
T_{r} = \frac{T_{sky}\frac{P_{on}}{P_{off}} - T_{ref}}{1 - \frac{P_{on}}{P_{off}}}
\end{equation}
Once $T_{r}$ is known from these measurements, we can find temperatures for other unknown sources by replacing $T_{sky}$ with the unknown $T_{source}$. Let $P_{source}$ be the power measured from the source and $T_{source}$ be the corresponding temperature. Then we can rewrite Eq.~(\ref{eq:Tr}) to obtain
\begin{equation}\label{eq:Tsource}
T_{source} = \frac{T_{r}(1 - \frac{P_{source}}{P_{off}}) + T_{ref}}{\frac{P_{source}}{P_{off}}}
\end{equation}

The antenna has a large beam and might pick up stray radiation from the surrounding structures or ground at lower elevation angles. This effect contributes heavily to the system temperature. Hence, it is recommended to measure $T_{sky}$ by pointing the antenna near the zenith. The system temperature is the sum of all the temperatures that contribute to the output of the receiver, that is,

\begin{figure}[h!]
\centering
\includegraphics[scale=0.8]{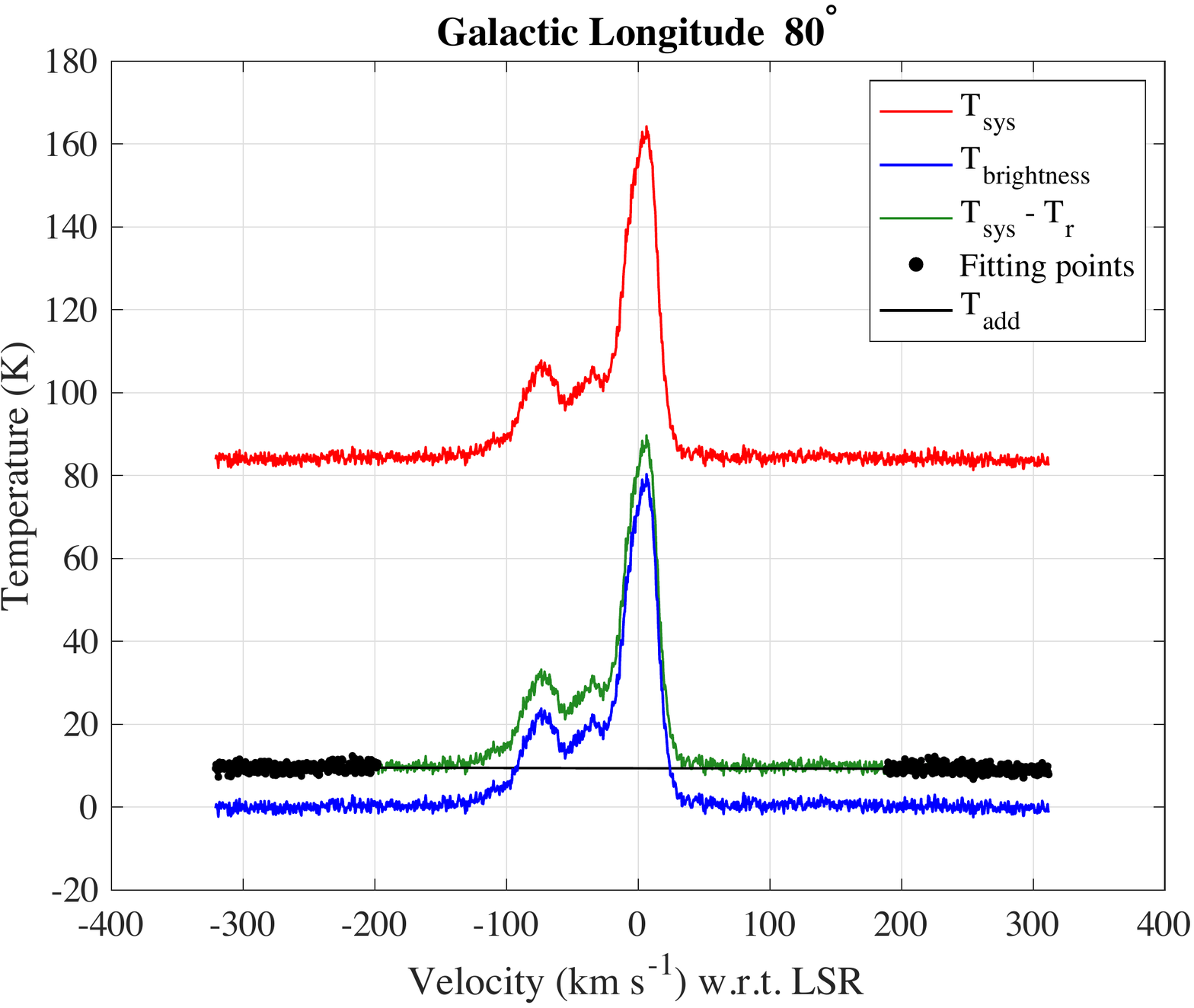}
\caption{(Color online) The plots show temperature calibration. The red curve indicates the system temperature. After subtracting the radiometer temperature from the red curve, we get the green curve. This still has some uncalibrated temperature ($T_{add}$) as per Eq.~(\ref{eq:Tsys}). It can be removed by fitting a line to the baseline of the green curve and subtracting it from the same curve. The black points and the black line represent this fitting. Finally, we get the blue curve, which is the brightness temperature of the source.}
\label{pcal}
\end{figure}

\begin{equation}\label{eq:Tsys}
T_{sys} = T_{source} + T_{r} + T_{sky} + T_{add}
\end{equation}
$T_{r}$ is usually the major term contributing to the system temperature. $T_{add}$ is an additional temperature, which is the sum of other temperatures that are not measured here. These can be the ground temperature picked up by the antenna through sidelobes, backlobe, or even frontlobe (especially for lower elevation observations), ambient temperature, gain fluctuations, etc. This term is depicted in Fig. \ref{pcal} as the green curve. Since our source is a spectral line source, we can remove this continuum temperature contribution by fitting a straight line to the baseline and subtracting it, which gives us the brightness temperature of the source, as shown in Fig. \ref{all_lines}. For continuum sources, each of these terms needs to be carefully measured and subtracted to determine the brightness temperature of the source.

The sensitivity of the measurement is given as
\begin{equation}\label{eq:sigma}
\sigma = T_{rms} = \frac{T_{sys}}{ \sqrt{\Delta f \Delta t}}
\end{equation}
where $\Delta f$ is the frequency bin size and $\Delta t$ is integration time. The integration time $\Delta t = \Delta t_{on} + \Delta t_{off}$. As the time is equally shared between on-source and off-source measurements for each observation, $\Delta t_{on} = \Delta t_{off}$. This is the hot-cold measurement method of calibration. Hence, Eq.~(\ref{eq:sigma}) can be rewritten as

\begin{equation}\label{eq:Trms}
T_{rms} = \frac{\sqrt{2}T_{sys}}{ \sqrt{\Delta f \Delta t}}
\end{equation}
For the observations presented here, the spectral resolution $\Delta$f  is 4 kHz and the integration time $\Delta$t is 4 sec. The value for $T_{rms}$ estimated using Eq.~(\ref{eq:Trms}) is 0.75 K, while the average $T_{rms}$ measured from observations is 0.9 K. This slight difference is because the gain fluctuations are faster than the time required for a single observation, which are not accounted for in Eq.~(\ref{eq:Trms}). Faster switching between a load and the source or fast frequency switching can remove the gain fluctuations. Another effect that contributes to larger values of measured root-mean-square (RMS) temperature is the fact that $T_{sys}$ is higher where the line is present in the spectrum. This increases $T_{rms}$ for frequencies for which the line is present in the spectrum, as seen in Fig. \ref{gauss}.

The calibration is cross-checked by pointing the antenna in the direction of the S8 region in the Orion constellation. We observe a good agreement between peak velocity in our measurement and that of Williams and Green.\cite{williams,Green} This is also verified by the Haystack Observatory Small Radio Telescope (SRT).\cite{Haystack} The peak brightness does not agree with these measurements because of the effect of the large beam of the antenna, as seen in Eq.~(\ref{eq:Ta}). The temperature $T_A$ detected by the antenna is

\begin{equation}\label{eq:Ta}
T_{A} = \frac{1}{4\pi} \int_{0}^{2\pi} \int_{0}^{\pi} G(\theta,\phi) T_{b}(\theta,\phi)\,d\theta d\phi
\end{equation}

where $G(\theta,\phi)$ is the antenna gain and $T_{b}(\theta,\phi)$ is the sky brightness temperature distribution. The shape of the line is similar to those observed in Green, Williams, and SRT. To understand the difference seen in the peak brightness temperature of the S8 region (and other sources observed here), we must consider the temperature distribution of the extended source $T_{b}(\theta,\phi)$. The brightness temperature seen by the antenna is then the temperature integrated over the beam of the antenna, as seen in Eq.~(\ref{eq:Ta}). Hence, different beam widths will detect different peak brightness temperatures for the same source.\cite{Green,williams,Haystack} This is basically due to the beam averaging of the source.

\begin{figure}[h!]
\centering
\includegraphics[scale=0.7]{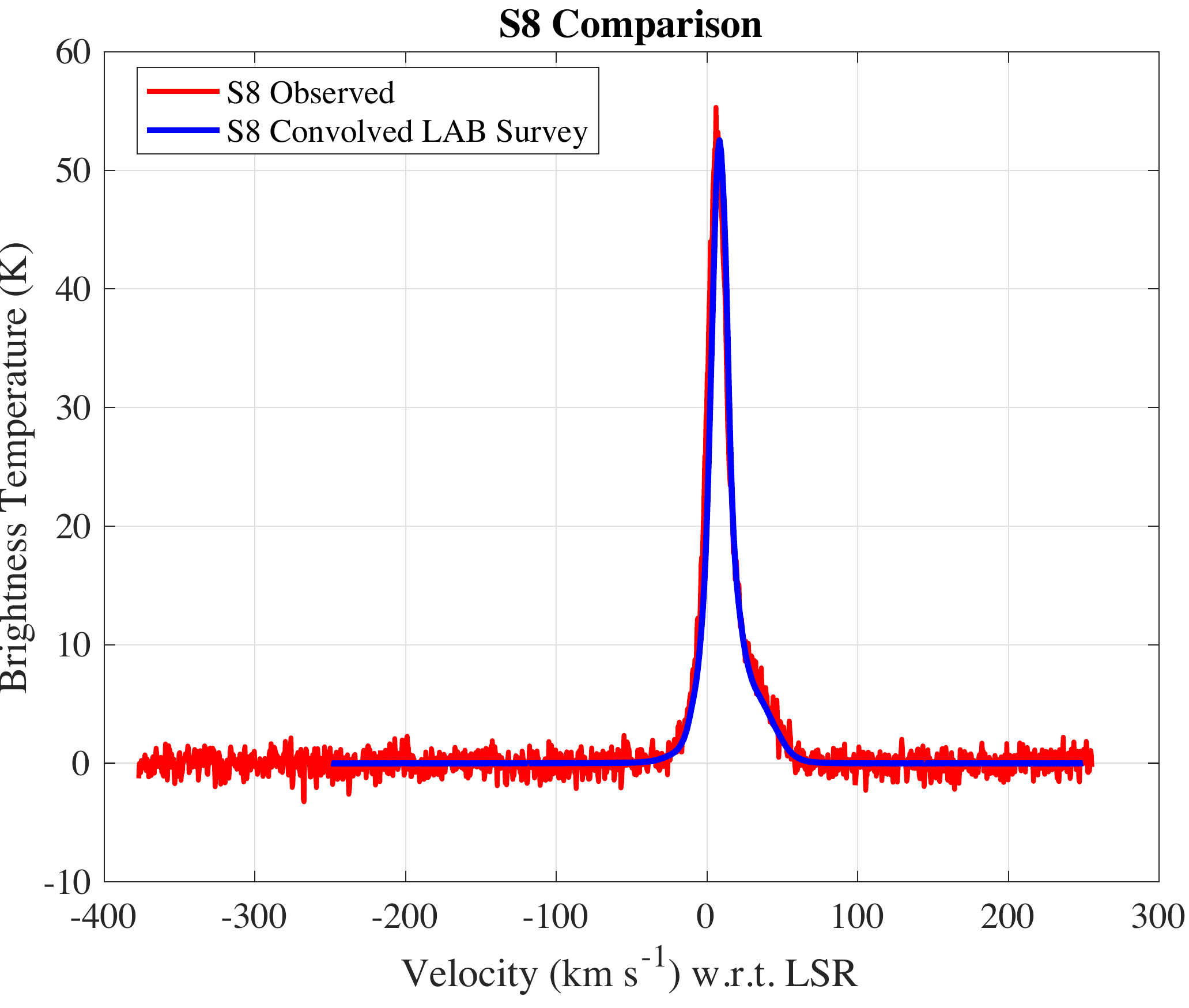}
\caption{Measurement of S8 region in Orion constellation. The observed S8 line profile is compared with the data from the LAB survey convolved  with the antenna beam.}
\label{S8}
\end{figure}

Observation is compared with the Leiden/Argentine/Bonn (LAB) survey of Galactic HI to verify the expected brightness temperature of the S8 region.\cite{LAB} For a true comparison, data from the LAB survey is convolved with a Gaussian of the size of the antenna beam, and the line profile for the S8 region is plotted along with the observed S8 region. As seen in Fig. \ref{S8}, observation is in excellent agreement with the convolved LAB survey HI line profile.

Finally, the  line velocity  (\emph{x}-axis) is converted from frequency to velocity using the Doppler effect equation\cite{Doppler} 

\begin{equation}\label{eq:v}
v = c\big(1- \frac{f}{f_0}\big)
\end{equation}
where $c$ is the speed of light, $f_0$ is the rest frequency of the hydrogen line, and $f$ is the received (Doppler-shifted) frequency. The frequency $f$ varies with
date and time of observations as well as the position of emitting source on the sky.
However, to derive the  velocity of the  21 cm line emitting  hydrogen cloud in the local rest frame (called the \emph{Local Standard of Rest} or LSR)
we need corrections  for  following three main factors that move the  telescope relative to the hydrogen cloud (note this speed  is not the  rotation  speed  around the
galactic center):
\vskip 0.2cm

(1)  The rotational or spin motion of Earth on its axis, and how fast the telescope rotates in space (which is a function of the observational location) 

(2) The orbital  motion of Earth around the Sun (average orbital speed is about 30 km/s, but due to eccentricity of Earth's orbit this speed changes slightly over the orbit) 

(3) The motion of the Sun  within the local group of stars (this is about  20 km/s) 

When we correct for  all  these three movements of the telescope we have finally placed ourselves in the  local reference system  or the Local Standard of Rest as defined above. 
The correction factor can be calculated using the Python script included in the Supplementary Materials. One can also use the online calculator provided by the Green Bank Telescope,\cite{GBT_rvc} but this calculator is valid only for the location of Green Bank Telescope. As a consequence, a small error arises depending on the location of the observing telescope on Earth. Our Python script takes into account this error and gives a more accurate result. After correcting the velocities to the LSR, we can obtain  the  rotation curve and  the  structure of the Milky Way galaxy.

\begin{figure}[h!]
\centering
\includegraphics[scale=0.44]{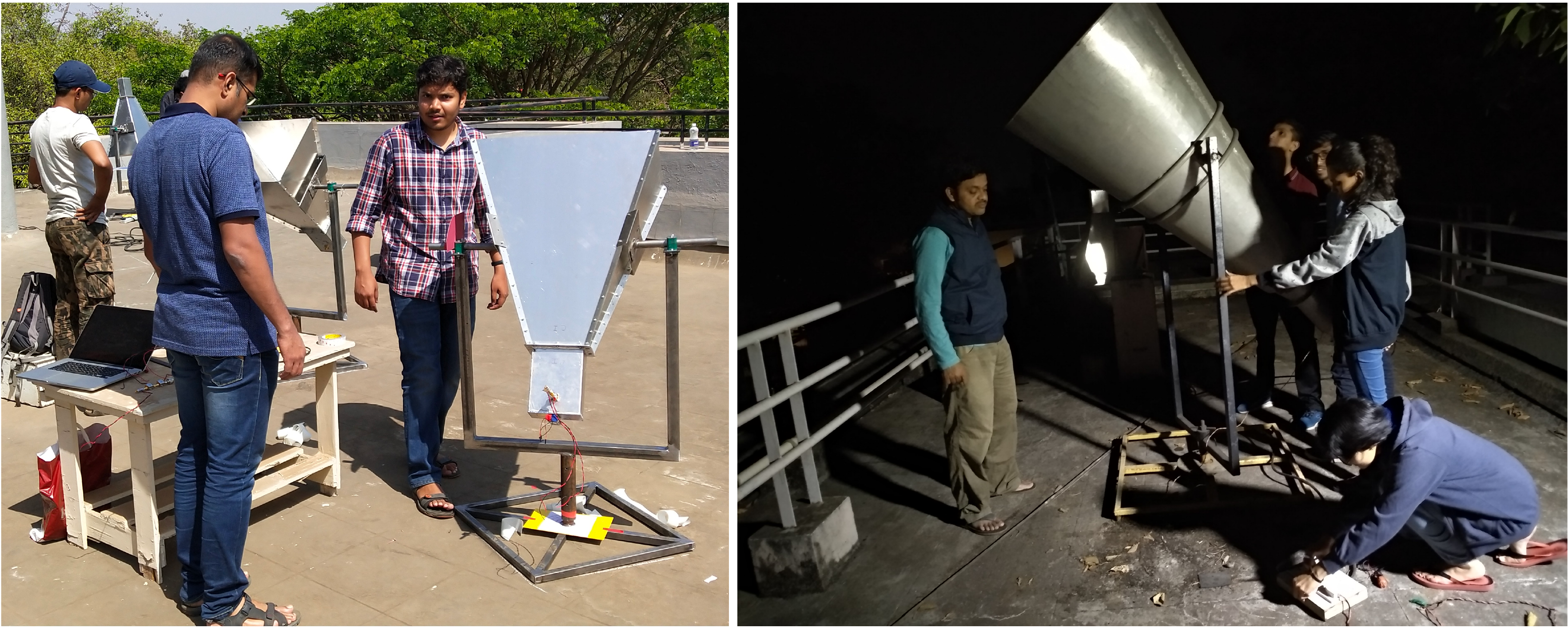}
\caption{Students performing hydrogen line experiments with the horns described in this paper 
at the Inter-University Centre for Astronomy and Astrophysics (IUCAA), Pune, India.}
\label{sessions}
\end{figure}

\subsection{A note on calibration} 

An alternate hot-cold calibration method uses two sources with known absolute temperatures. These are usually liquid nitrogen or liquid helium in standard experiments. We have avoided using these low-temperature components due to their complexity and difficulty in assembling.

In the calibration method we described in the previous section, a microwave absorber should ideally be placed on the wall or directly in front of the horn opening for the on-source measurement during calibration. This should avoid reflections and make the antenna temperature equal to the absorber. Our calibration shows that an absorber may not be strictly required for such a setup if the wall is in the far-field of the antenna and the antenna is not pointing perpendicular to it. However, calibration errors may arise due to reflections from the wall without an absorber. For a sanity check, one should always compare results with a standard survey, as done in Fig. \ref{S8}.

Another way to calibrate the system would be to compare a HI region with the LAB survey by convolving the antenna beam with the data from the survey as in Fig. \ref{S8}. $P_{on}$ and $P_{off}$ in Eq.~(\ref{eq:Tr}) will correspond to peak temperature and baseline temperature, which is zero, of the HI region from the survey. That is, we convert the data to temperature-velocity plot, remove the offset and scale the peak temperature to match it to the convolved LAB survey plot. This scaling factor can then be used to calibrate other plots.

\section{Results}

\begin{figure}[h!]
\centering
\includegraphics[scale=0.75]{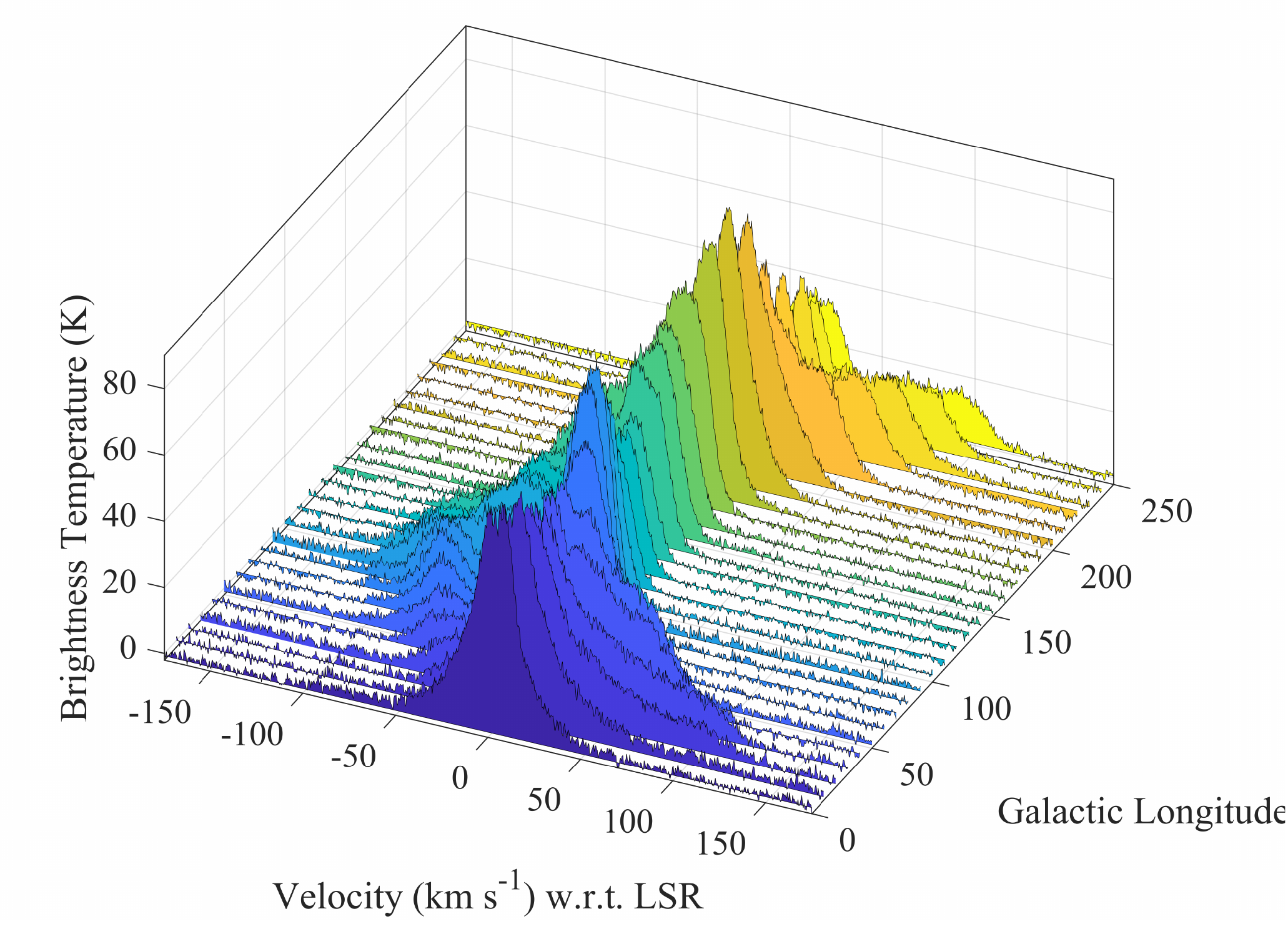}
\caption{All observations taken for Galactic longitudes (L) $0 \leq L \leq 250$ degrees at intervals of 10 degrees.}
\label{all_lines}
\end{figure}

Figure~\ref{all_lines} displays plots of the hydrogen line taken at 26 different equally spaced Galactic longitudes. We note that in all of the plots the hydrogen line is broad, and there are multiple peaks distinctly visible in some plots. Spectral lines have a finite width due to the Doppler broadening of the line.\cite{ra_book, plasma,Doppler} The atoms of neutral hydrogen have a Maxwellian velocity distribution due to thermal agitation. The thermodynamic temperature of the neutral hydrogen gas is encoded in the width of a spectral line. The typical temperature of cold neutral hydrogen is around 100 K, which would give a line width of about 10 kHz. However, the observed line width is much larger than this predicted value because the measured line profile is integrated over several small Doppler-shifted hydrogen lines that the telescope cannot resolve. This can be attributed to contributions from the warm neutral medium, which has temperatures from several hundred to thousand degrees.\cite{Dickey_Lockman} A part of broadening is also associated with the turbulence in the hydrogen clouds.\cite{ra_book} Multiple peaks appear because our Galaxy is rotating, and the line of sight of the antenna crosses different arms of the Galaxy. Each point of intersection between a line of sight and a Galactic arm has a component of velocity along the line of sight. This component is responsible for the Doppler shift of the line corresponding to each arm in the line of sight. It is difficult to isolate peaks in some plots. This limit is partially imposed by the noise in the receiver and the large beam of the antenna.

\subsection{Fitting Gaussians}

\begin{figure*}[h!]
\centering
\includegraphics[scale=0.7]{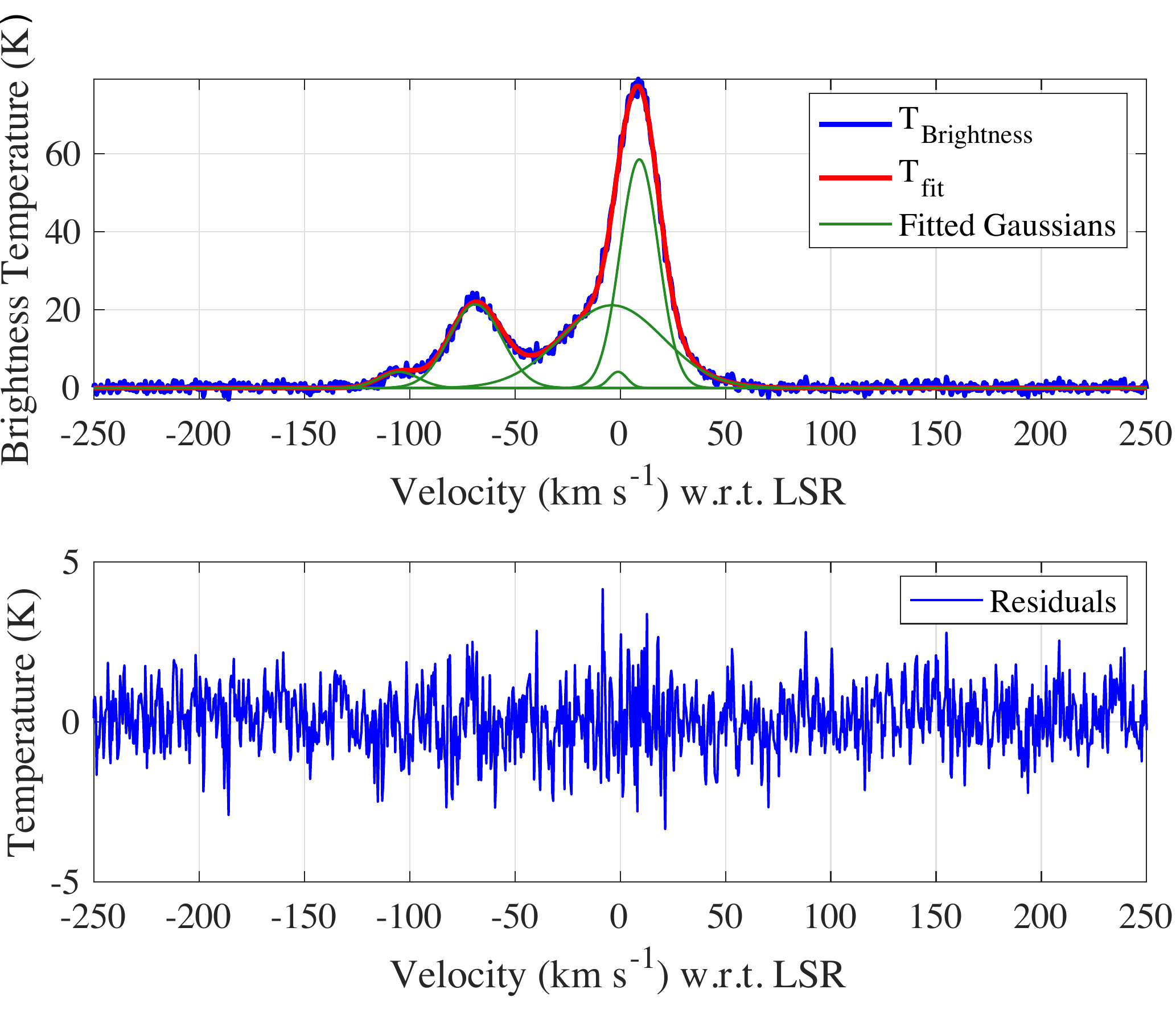}
\caption{The model obtained by fitting Gaussians, and the corresponding residuals (bottom panel) for the Galactic longitude 70 degrees.}
\label{gauss}
\end{figure*}

Gaussian functions are fit to each observation using the MATLAB fitting tool. The number of Gaussians is increased gradually until minimum RMS is seen in residuals. The sum of Gaussians is treated as the model. As discussed earlier, residuals seen in Fig. \ref{gauss} are larger at the peaks. This is because $T_{rms}$ depends on the $T_{sys}$. The system temperature at the peaks is higher as compared to the baseline, which gives higher $T_{rms}$ at the peaks. This also affects the fitting as some features of the line are buried in the residuals. 

\subsection{Rotation curve of the Milky Way Galaxy}

\begin{figure*}[h!]
\centering
\includegraphics[scale=0.33]{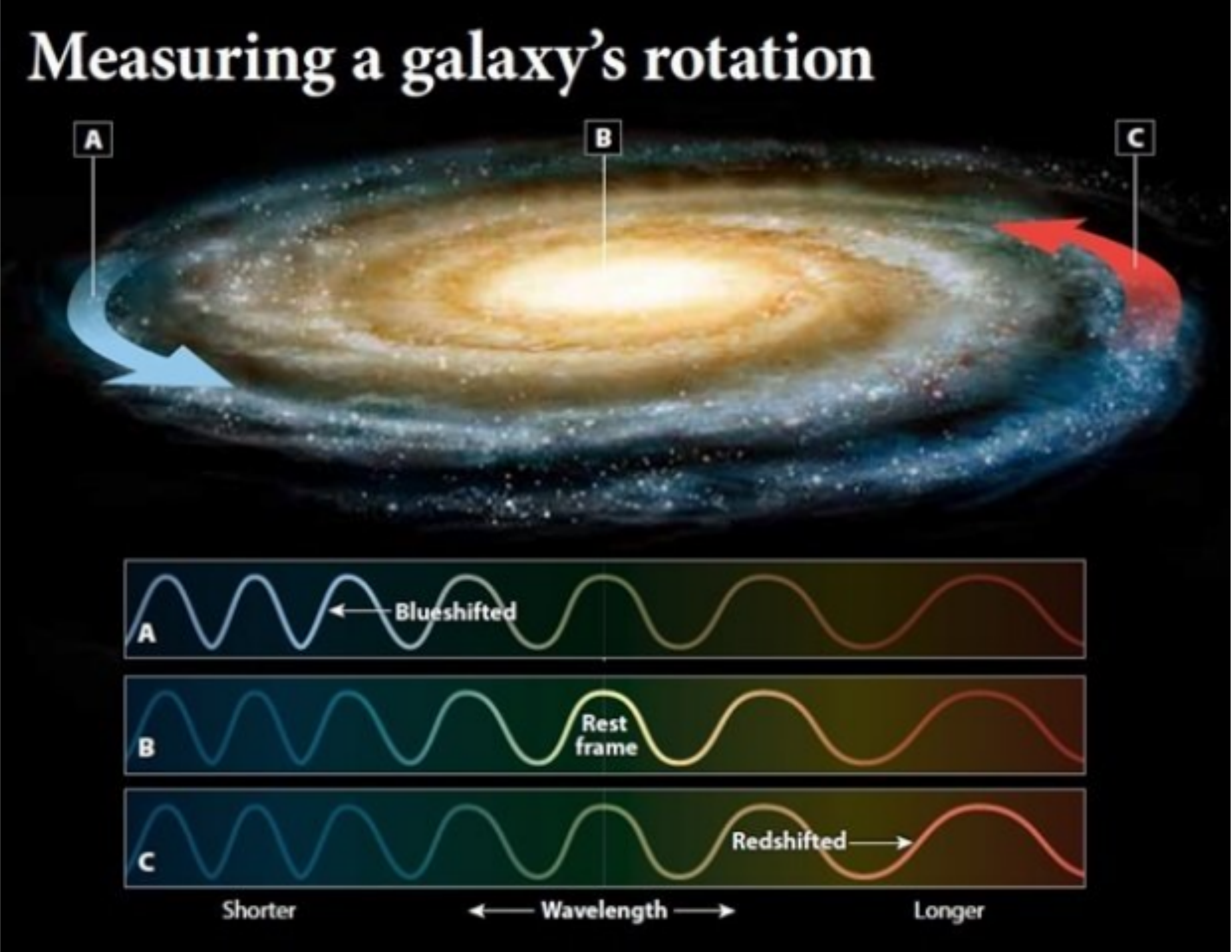}
\includegraphics[scale=0.41]{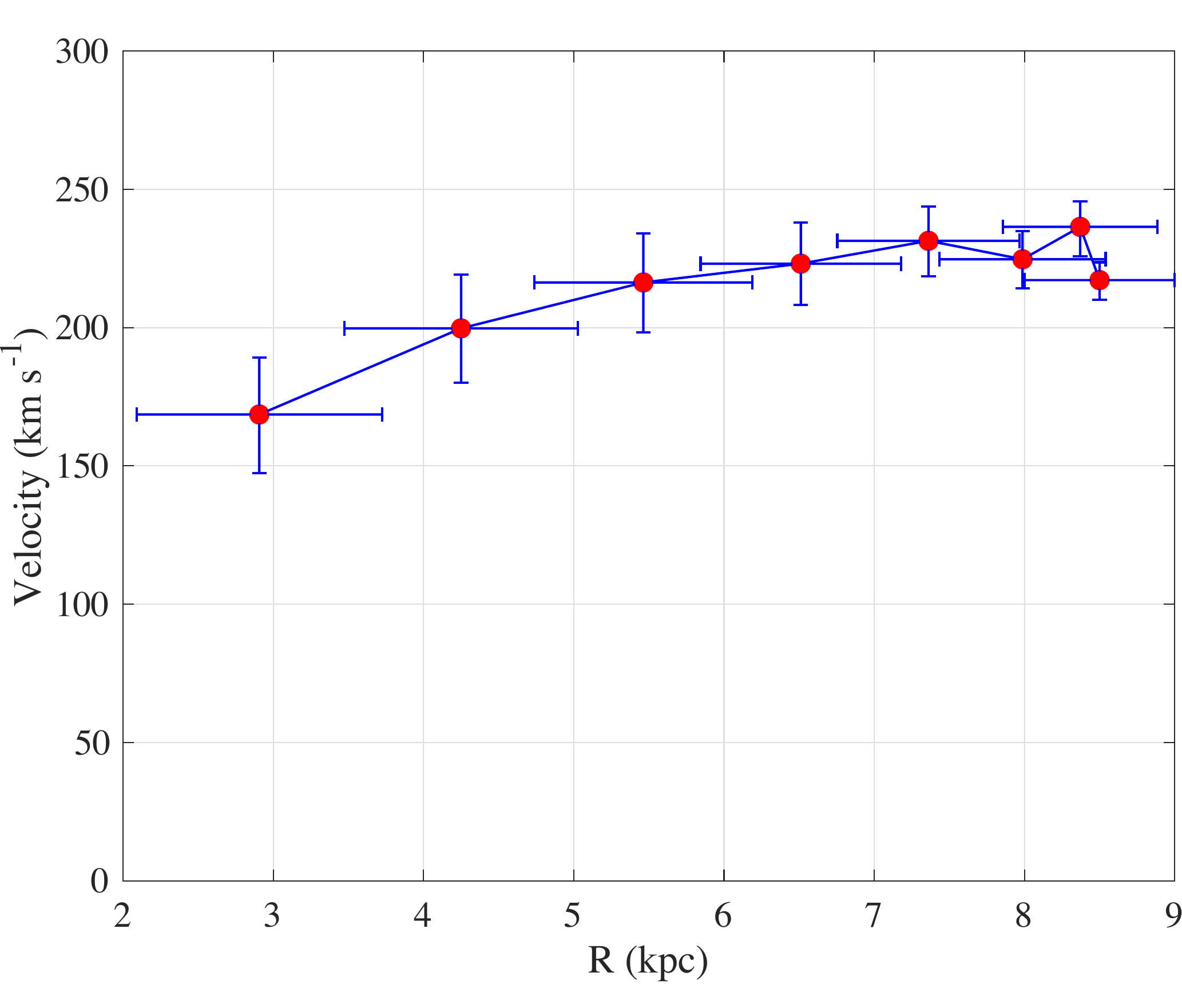}
\caption{Left: Schematic shows how one  can use  Doppler shift\cite{Doppler} of 21-cm hydrogen spectral line to infer the rotation
curve of a galaxy like the Milky Way. Hydrogen clouds at points A and C will show  maximum blue and red shifts  from the rest wavelength while point B will emit  at the  
rest wavelength (Image Credit: Physics OpenLab). Right: This figure  shows the  rotation speed of the Milky Way galaxy  obtained  as a function of distance 
$R$ from galactic center, from our observations. The Sun is located at about 8.5 kpc from the center of our Galaxy. A flat rotation curve is observed at this radius with an average rotation speed of  218 km/s. This graph proves that our Galaxy is rotating  but also that it rotates differentially, not as a solid body but more like a fluid. From such curves astronomers can infer the  presence of mysterious dark matter in spiral galaxies.\cite{Corbelli}}
\label{rot_curve}
\end{figure*}

Before 21 cm hydrogen line observations, we did not know  well how our Galaxy rotates; whether it spins like a solid disc, or like planets orbiting around the Sun following the Kepler's laws or in some other way. To get an idea how this can be done, we point our telescope to various galactic longitudes  L  and  obtain the rotation curve of the Milky Way galaxy from observation  on the Galactic longitudes $0 < L < 90$ degrees using the following equations:
\begin{equation}
V = V_{r} + V_{\odot} \times \sin L  
\end{equation}
\begin{equation}
R = R_{\odot} \times \sin L
\end{equation}

where $V_{r}$ is the most red-shifted velocity in the plot. $V_{\odot}$ and $R_{\odot}$ are the rotation speed of the Sun and the distance of the Sun with respect to the Galactic center, respectively. $V_{\odot}$ and $R_{\odot}$ here are considered to be $218 \pm 6$ km/s and $8.5 \pm 0.5$ kpc, respectively. \cite{Bovy} Choosing the most red-shifted point is limited by the noise in the system. These points are chosen from the model by giving a $3 \sigma$ cutoff. The error bars on this point are given by $\pm 1 \sigma$ around it. Average recession velocity was calculated from $L=180$ degrees Galactic longitude observation,\cite{Liu} and it was found to be $25 \pm 1$ km/s. This is subtracted from the maximum redshift-velocity to account for line broadening due to the thermal and turbulence effects. The noise increases when pointing towards the direction of the Galactic center. Also, the line profiles are much broader for $L<30$ degrees, and they become narrower for $50<L<90$ degrees. As the measurements are limited by noise, choosing the most red-shifted points is affected by noise for lower Galactic longitudes. This accounts for the deviation of the rotation curve at lower Galactic longitudes (see Burton,\cite{burton} for comparison).

Figure~\ref{rot_curve} shows the derivation of rotation curve of the Milky Way galaxy plotted up to the solar radius (Sun is located about 8.5 kpc from the center of our Galaxy) from our observations. 
At Sun's position  a flat rotation curve is observed  with an average rotation speed of  $218 \pm 6$ km/s.  Detailed observation reveal that this flat rotation extends even beyond the orbit of the Sun and and that it is present in most other spiral galaxies (like the Andromeda galaxy). The expected rotation speed of the Galaxy near the Sun is only about 160 km/s, if we consider the gravity of all the visible matter. Therefore, this plot proves that not only is our Galaxy rotating faster than expected, but also that it rotates differentially, i.e., it rotates not as a solid body but more like a fluid.  To account for the flat rotation curves, astronomers have invoked  the  presence of  a mysterious  dark matter in spiral galaxies and also in the universe.\cite{Corbelli}

\subsection{Structure of the Milky Way Galaxy}

\begin{figure}[h!]
\centering
\includegraphics[scale=0.45]{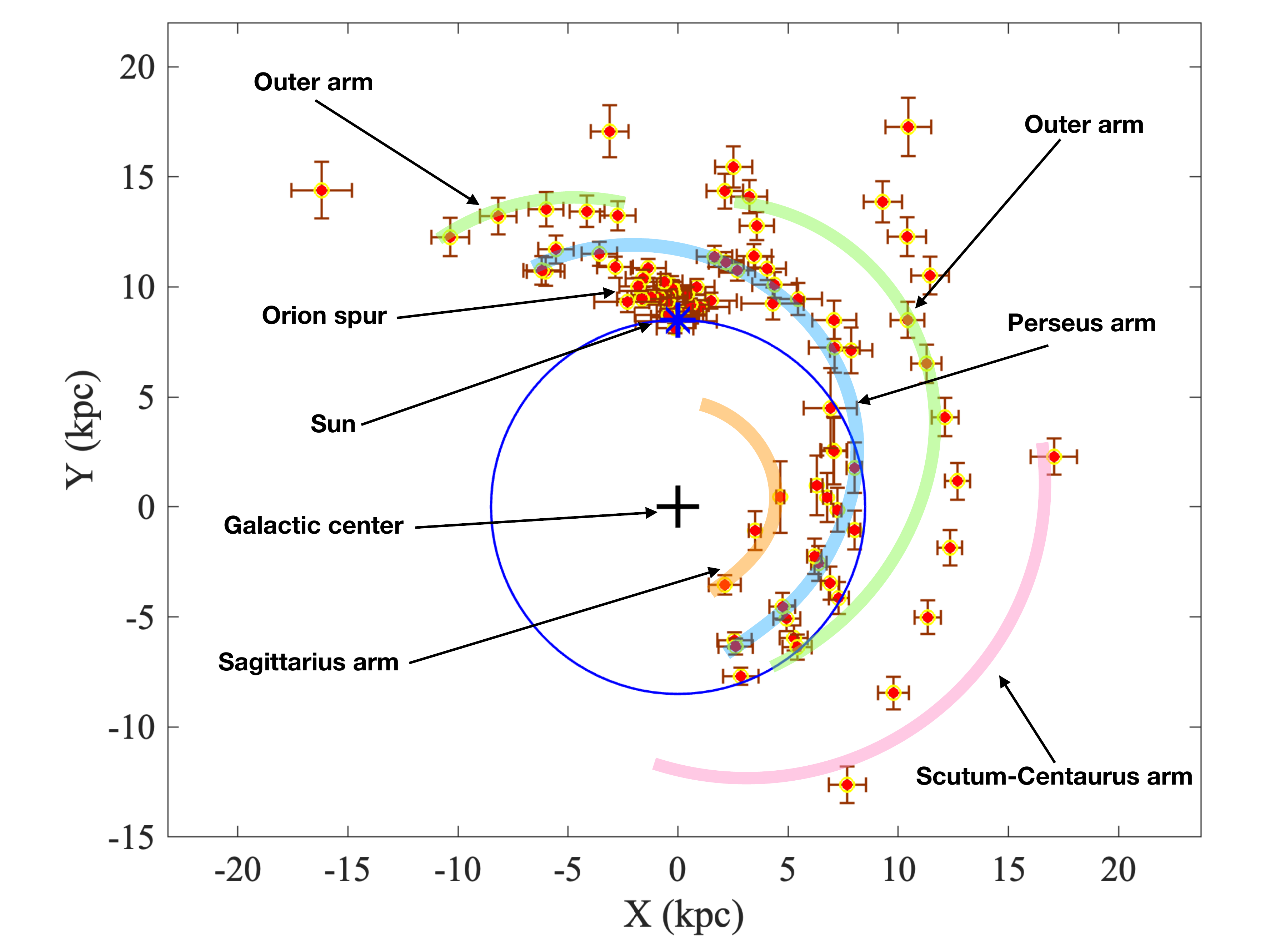}
\caption{(Color online) The structure of the Milky Way galaxy obtained using our hydrogen line observations. The blue star represents the Sun, and the blue circle is the solar orbit in the Galaxy. Each point represents one peak of an observation. The spiral arms of the Galaxy (plotted in color) can be seen distinctly.}
\label{structure}
\end{figure}

\begin{figure}[h!]
\centering
\includegraphics[scale=0.8]{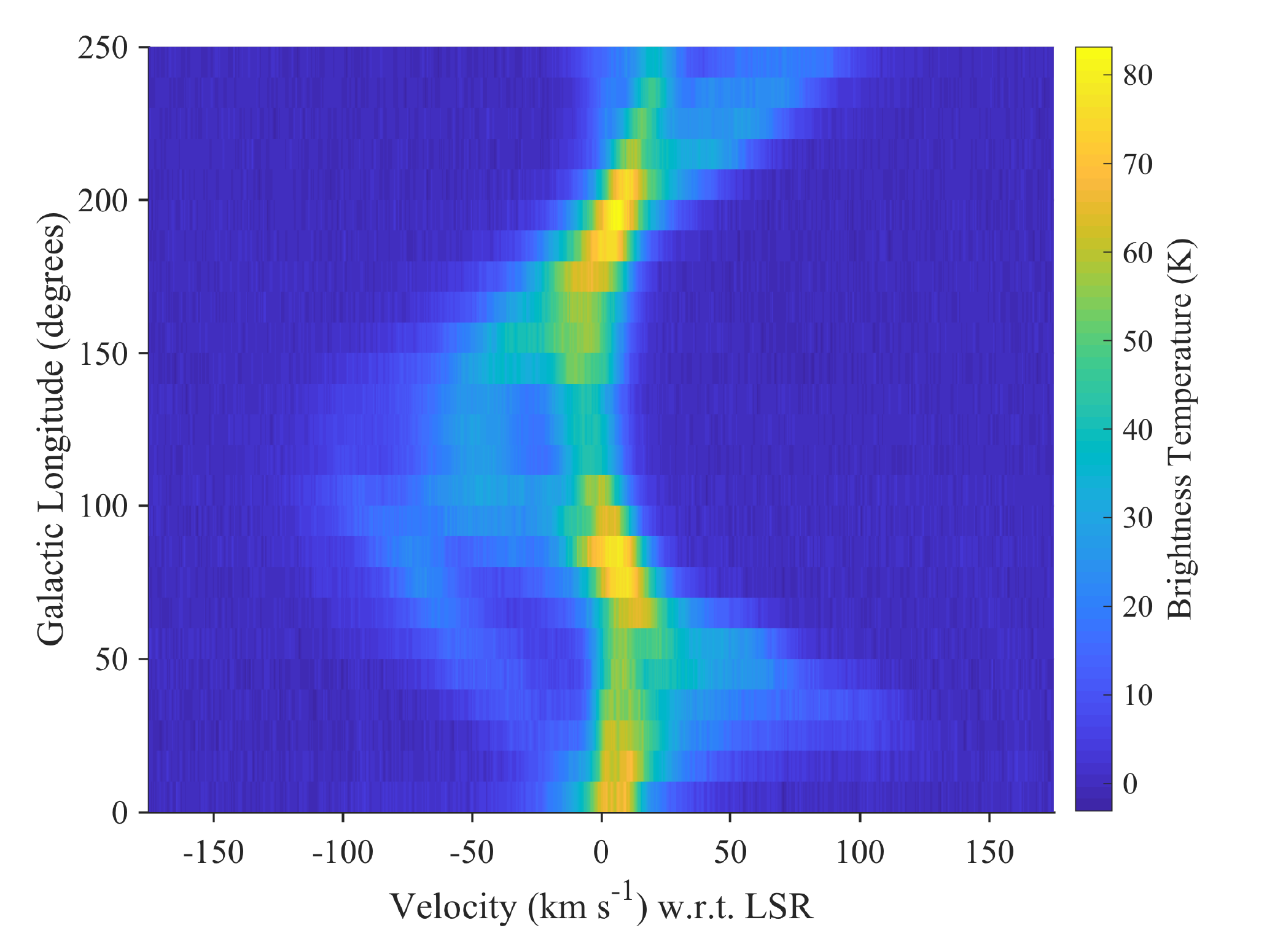}
\caption{The figure shows the longitude-velocity plot of the Milky Way galaxy compiled using our observations. The neutral hydrogen gas distributed in spiral arms form distinct curves in the plot.}
\label{2D}
\end{figure}

\begin{figure}[h!]
\centering
\includegraphics[scale=0.5]{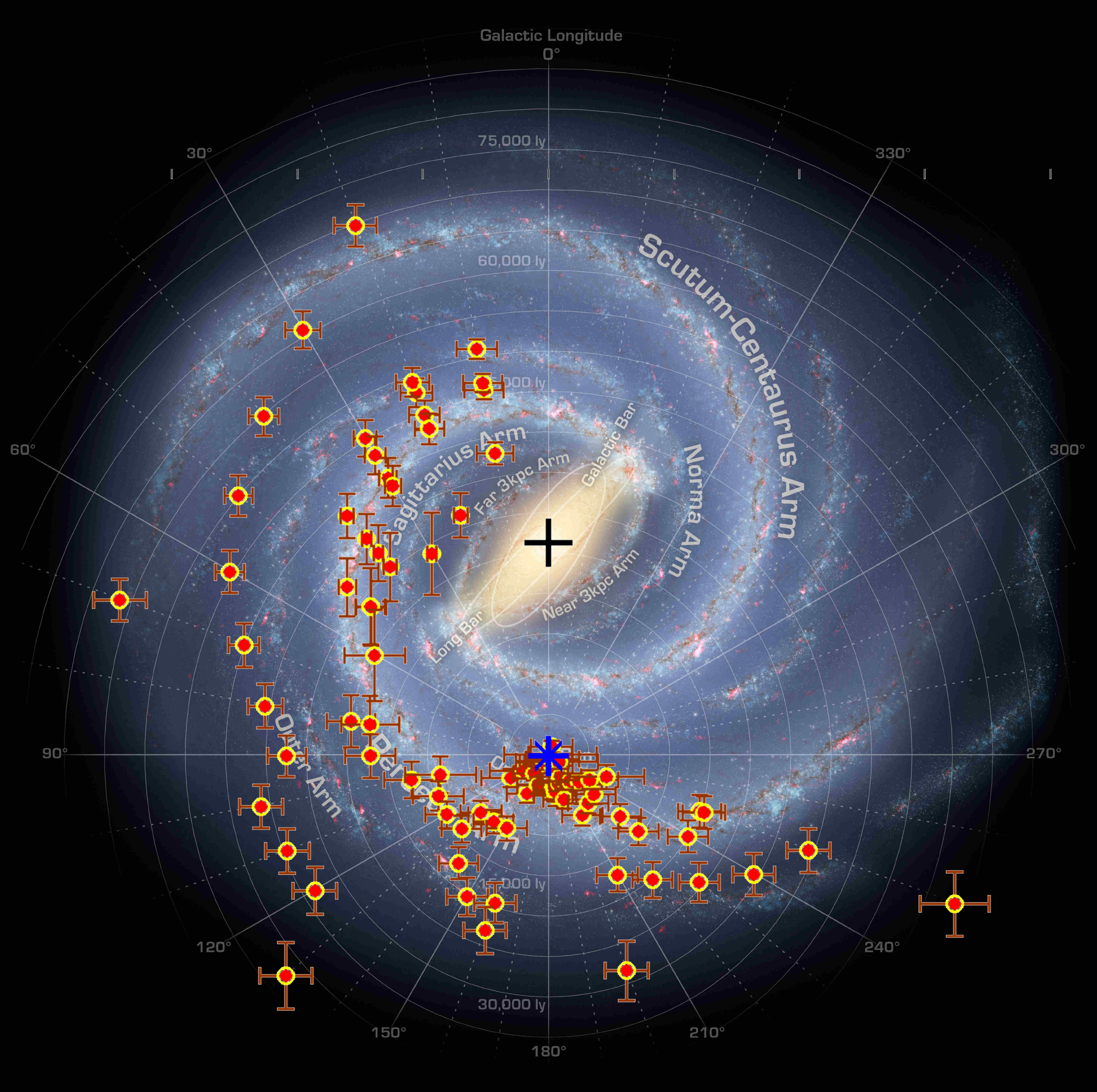}
\caption{(Color online) A comparison of the structure of the Milky Way galaxy obtained by the experiment presented here along with the artistic image produced by NASA based on various surveys.\cite{NASA} The blue star is the Sun.}
\label{comparison}
\end{figure}

The Galactic structure can be derived using the observations given in Appendix A. This requires knowledge of peaks of fitted Gaussians and the rotation curve, which we have obtained in Fig.~\ref{rot_curve}. The position of each peak can be plotted on a polar plot in terms of $R$ and $\theta$. The details of the method can be found in Hulst and Liu.\cite{hulst,Liu} Errors on parameters were obtained from Bovy \emph{et al.}\cite{Bovy} For sources within the solar orbit, there is an ambiguity in determining the positions of the peak. The line of sight intersects circles within the solar orbit at two points. As a consequence, both these points produce the same Doppler shift.\cite{Doppler} Hence, 21 cm line peaks from within solar orbit can have two possible locations in Fig. \ref{structure}. Only one point is plotted for these directions for clarity. Figure~\ref{2D} is the longitude-velocity plot of the Galactic plane. Another method to derive the Galactic structur
e is by recreating the longitude-velocity plot from spiral models.\cite{simonson,burton_density_wave} We see in Fig.~\ref{comparison} that the Outer arm merges with the Scutum-Centaurus arm. This effect occurs because some of the peaks are not resolved due to the noise in the measurements and the large beam of the antenna. The measured line profile is the sum of several overlapping peaks. 

A major component in the uncertainty in measurement is the noise in the receiver. This can be reduced by using low noise components, sensitive analog-to-digital converters, and cooling the front-end electronics to a very low temperature. Such resources are not always available to the students. A simple alternative is to have a longer integration time.

\section {Other 21 cm line projects}
It is worthwhile to compare BHARAT with other similar projects. Radio Jove\cite{radio_jove} is one such notable project that focuses on radio emissions from Jupiter and the Sun. The other two prominent projects, which are primarily designed around the detection of the 21 cm hydrogen line, are the Small Radio Telescope (SRT) (also mentioned earlier) and Digital Signal Processing in Radio Astronomy (DSPIRA).\cite{SRT, DSPIRA}

SRT was developed by the Haystack Observatory, USA, which uses a motorized dish antenna. The cost of the setup is between 7k -- 9k USD, and the sensitivities are between 150 K to 170 K. One advantage of SRT over BHARAT is the narrow beam of the antenna, which can give a more accurate measurement of the galactic structure.

The DSPIRA project was developed at the West Virginia University, USA. It has a simple horn antenna assembly and other options for LNAs and filters, which can be easily incorporated here as well (if desired). Both projects have nearly similar costs. The DSPIRA project is developed for digital signal processing (DSP) in radio astronomy, while  BHARAT focuses more on the instrumentation and astrophysics part of the experiment. However, it can also be designed for DSP experiments by simple software modifications.

\section{Conclusion}
We have designed a cost-effective radio telescope for use in laboratory training in radio astronomy at colleges and universities. The system employs a simple dual-mode antenna and off-the-shelf electronic components. The antenna is designed for low sidelobes, thereby reducing system noise significantly. The noise $T_{rms}$ is about 0.9 K with an integration time of 4 s. The results presented illustrate the capability of this radio telescope. We have demonstrated that this setup can produce significant results and help students understand radio astronomy techniques and the basic concepts in astrophysics. It is being used for experiments and demonstrations at our lab and  universities across the country. It has also motivated several students to take up experimental radio astronomy as their curricular and extra-curricular projects and helped them to select astronomy
 and astrophysics as their career option. With the advent of next generation radio telescopes like Square Kilometer Array (SKA),\cite{ska} 
 Low Frequency Array (LOFAR),\cite{lofar} and Long Wavelength Array (LWA),\cite{lwa} it is imperative that students,  as aspiring researchers, acquire a good understanding of radio astronomy science.   The simple antenna design presented here has adequate sensitivity and broad antenna beam width, ideally suitable for teaching fundamentals of radio astronomy in colleges at low cost. In addition,  this setup can be  modified in the future by  adding more antenna elements in a  phased-array, multi-beam concept, using cryogenic cooled low-noise front-end, a better  back-stage detection circuit, and employing a wider tuning range with advanced signal processing  software. This will make the telescope far more versatile and sensitive to faint, and fast varying cosmic signals. We envisage this capability  will  be  of  use in many  areas like pulsar detection,  Hydroxyl  (OH) Megamaser line study,\cite{megamaser} Search for Extra Terrestrial Intelligence (SETI),\cite{seti} and in tr
ansient detection experiments (similar to STARE2).\cite{STARE2} Ultimately, how such telescopes can be used in  radio astronomy is only limited by the ingenuity, imagination, and skill of the user.
 
 \newpage
 \section{Acknowledgments}

 We thank the workshop department of the National Centre for Radio Astrophysics - Tata Institute of Fundamental Research (NCRA-TIFR) for constructing the antenna. We thank Prof. Avinash Deshpande for his suggestions. Special thanks to Shishir Sankhyayan for assisting with the MATLAB codes and Pratik Dabhade for his valuable inputs regarding astrophysics. We thank Prof. Kelvin Bandura for his correspondence on DSPIRA related information. We would also like to thank Jameer Manur and the students at the Radio Physics Laboratory for their timely help in managing the logistics. JB thanks Christ (deemed to be) University and IUCAA for all the support received. JJ would like to place on record the support he received from IUCAA in the form of the Visiting Associateship.



\section{Appendix A: Measurements in Galactic plane}\label{appendixa}

\begin{figure*}[htbp!]

\includegraphics[scale=0.75]{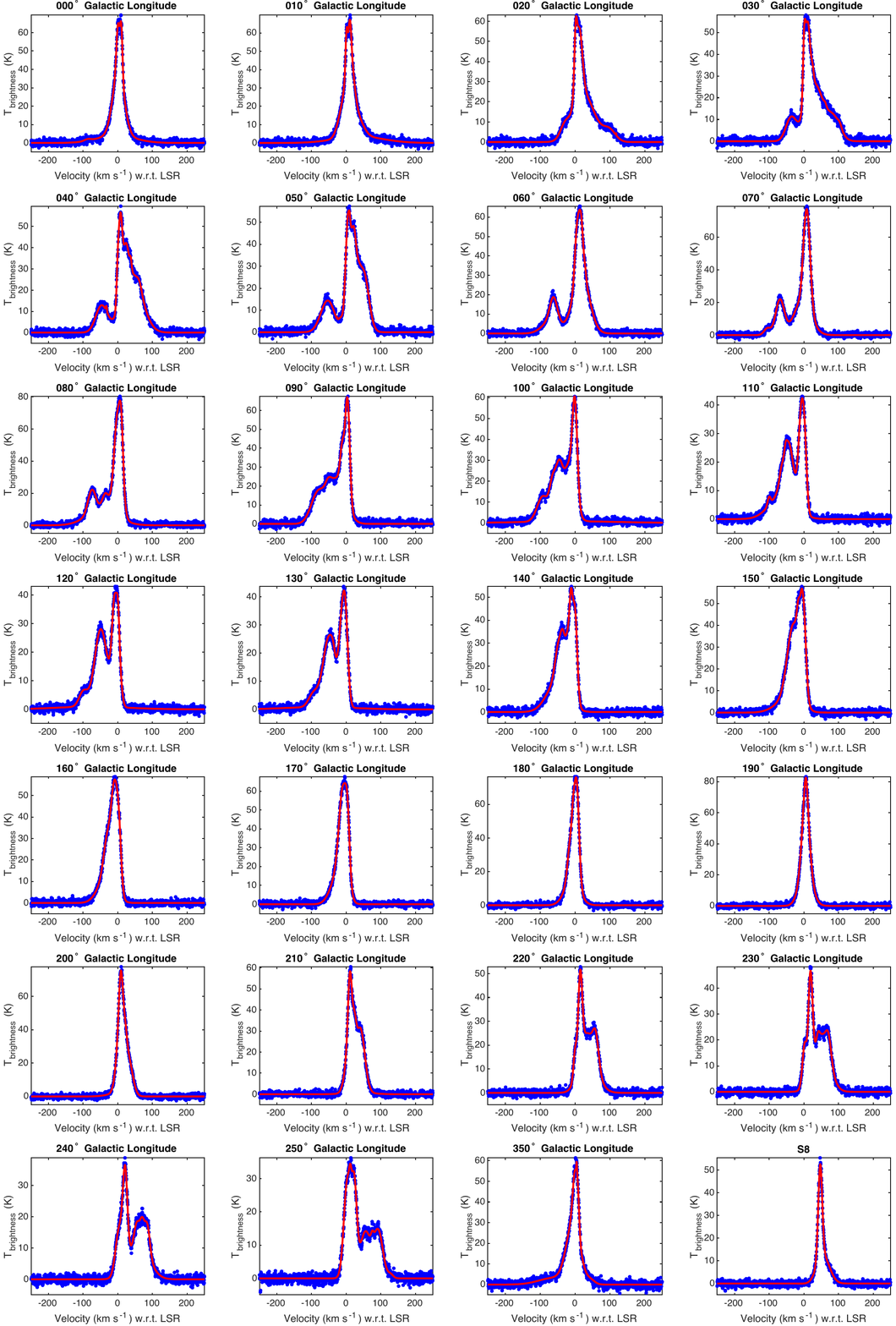}
\caption{A collage of 21 cm line profiles measured in Galactic plane. The galactic longitude is given on top of each figure.
\emph{x}-axis denotes line velocity with respect to Local Standard
of Rest in km/s and \emph{y}-axis is brightness temperature
in Kelvin degrees.}
\label{plots}
\end{figure*}

Plots for all the measurements done in the Galactic plane are presented in Fig. \ref{plots}. Blue curves are data, and red curves are fit curves. The data can be found in Ref.~\onlinecite{data}.

\section{Appendix B: Pyramidal Horn}\label{appendixb}

\begin{figure*}[ht!]

\includegraphics[scale=0.4]{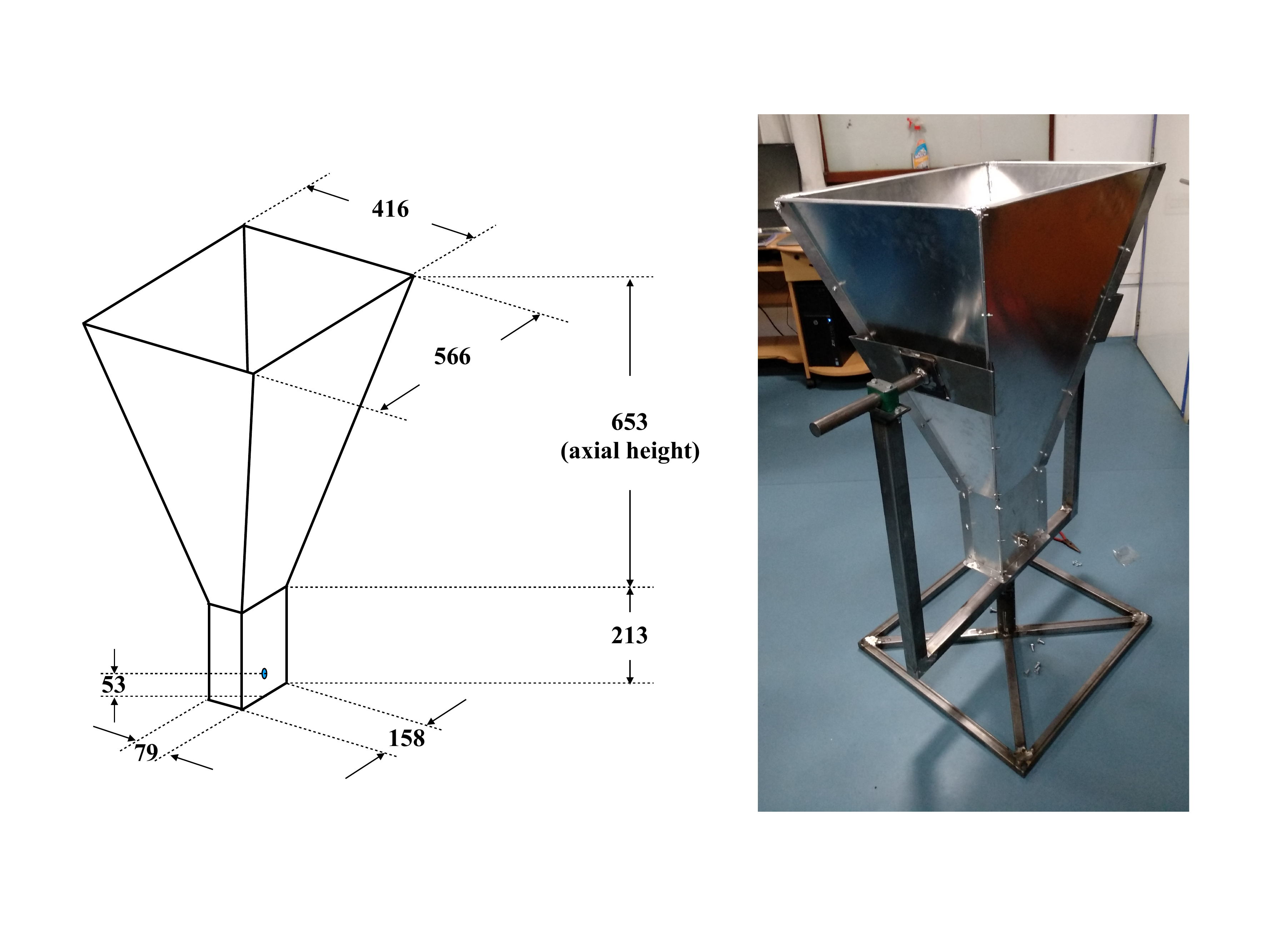}
\caption{Sketch of a simple Pyramidal horn with dimensions (in mm) and  photograph of the fully fabricated antenna. The antenna is fed through an N-type coaxial connector and is mounted on an altitude-azimuth mount. }
\label{pyramidal_horn}
\end{figure*}

The setup discussed here refers to the simple pyramidal horn described in Table~\ref{table}. The design and the picture of the antenna are shown in Fig. \ref{pyramidal_horn}. This setup is easier to assemble than a dual-mode horn; however, it comes at the expense of  lower gain, higher sidelobe level and larger beam. The gain of the antenna is 17 dBi, the sidelobe level is 12 dB, and the beam width (full width half maximum) is about  $26^{\circ}$. The radiometer temperature for the pyramidal horn setup at calibration was 144 K.

\end{document}